\documentclass[aip,jcp,amsmath,amssymb,reprint,numeric]{revtex4-1}

\usepackage{graphicx}
\usepackage{subfigure}
\usepackage{dcolumn}
\usepackage{bm}
\usepackage{mathrsfs}
\usepackage[dvipsnames]{xcolor}
\usepackage[normalem]{ulem}
\usepackage{caption}
\begin{document}
\title{Effective structure of a system with continuous polydispersity}
\author{Palak Patel}
\address{\textit{Polymer Science and Engineering Division, CSIR-National Chemical Laboratory, Pune-411008, India}}
\affiliation{\textit{Academy of Scientific and Innovative Research (AcSIR), Ghaziabad 201002, India}}
\author{Manoj Kumar Nandi}
\address{\textit{Polymer Science and Engineering Division, CSIR-National Chemical Laboratory, Pune-411008, India}}
\author{Ujjwal Kumar Nandi}
\address{\textit{Polymer Science and Engineering Division, CSIR-National Chemical Laboratory, Pune-411008, India}}
\affiliation{\textit{Academy of Scientific and Innovative Research (AcSIR), Ghaziabad 201002, India}}
\author{Sarika Maitra Bhattacharyya}
\email{mb.sarika@ncl.res.in}
\address{\textit{Polymer Science and Engineering Division, CSIR-National Chemical Laboratory, Pune-411008, India}}
\affiliation{\textit{Academy of Scientific and Innovative Research (AcSIR), Ghaziabad 201002, India}}

\begin{abstract}

In a system of N particles, with continuous size polydispersity there exists N(N-1) number of partial structure factors making it analytically less tractable. A common practice is to treat the system as an effective one component system which is known to exhibit an artificial softening of the structure. The aim of this study is to describe the system in terms of $M$ pseudo species such that we can avoid this artificial softening but at the same time have a value of $M<<N$. We use potential energy and pair excess entropy to estimate an optimum number of species, $M_{0}$.We then define the maximum width of the polydispersity, $\Delta \sigma_{0}$ that can be treated as a monodisperse system. We show that $M_{0}$ depends on the degree and type of polydispersity and also on the nature of the interaction potential,  whereas, $\Delta\sigma_{0}$ weakly depends on the type of the polydispersity, but shows a stronger dependence on the type of interaction potential. Systems with softer interaction potential have a higher tolerance with respect to polydispersity.  Interestingly, $M_{0}$ is independent of system size, making this study more relevant for bigger systems. Our study reveals that even $1\%$ polydispersity cannot be treated as an effective monodisperse system. Thus while studying the role of polydispersity by using the structure of an effective one component system care must be taken in decoupling the role of polydispersity from that of the artificial softening of the structure.

\end{abstract}

\maketitle
\section{Introduction}
Most systems that can be found in nature are inherently polydisperse. Polydispersity can be of different kinds like in size, in mass, and also in the shape of the particles. Also, the type of polydispersity and the degree of it varies with systems. 
Polydispersity brings variation in the properties of the material and  there are specially designed controlled experiments to create monodisperse particles\cite{exp_1_by_asher,Nucleation_in_Phase_Transitions}.
However, in some cases polydispersity is a desirable property. The size polydispersity is one of the most common types and it has been found that systems beyond certain value of polydispersity, known as the terminal polydispersity are good glass former
\cite{kofke1999,lacks1999,williams2001,pinaki2005,solid_liquid_transition_by_bagchi,Bidisperse_simulation,PDI_bagchi_sneha_sarika}.
It was shown that in a polydisperse system due to an increase in surface free energy, the crystal nucleation is suppressed promoting glass formation \cite{auer}. Thus in study of supercooled liquids, polydisperse systems play an important role.

In recent time, it has been shown that structure plays an important role in the dynamics of glass forming supercooled liquids \cite{role_pair_configuration,unraveling,effect_of_total_pair,validity_rosen,onset_crosspoint,role_pair_correlation,gcm_manoj,indranil,softness_manoj}. 
Since polydisperse systems are good glass former describing the structure of these systems becomes important. For a continuous polydisperse system, the number of species is the number of particles in the system. In this case, describing the system's partial structure in terms of independent species becomes an impossible task.
Thus it is common practice to treat a polydisperse system in terms of an effective one component system\cite{influence_PDI,colloidal_interaction_by_tanaka,PDI_in_colloids_by_salgi,role_of_PDI_michael,interfacial_PDI_michael} . However, it has been shown that we not only lose a large deal of information of the system by pre averaging the structure, the properties of the liquid thus predicted can also give spurious results. \cite{main_paper_Truskett,voigtmann_polydis_mct}. Truskett and coworkers\cite{main_paper_Truskett} have shown that for moderate polydispersity the thermodynamic quantities like the pair excess entropy obtained from the effective one component radial distribution function (rdf) predicts that with an increase in interaction the static correlation becomes weaker thus predicting structural anomaly. The study showed that when the system is expressed in terms of 60 pseudo neighbors and the excess entropy is calculated in terms of the partial structure factors (radial distribution functions), this structural anomaly disappears. Weysser {\it et. al.} while working with Mode coupling theory have shown that for a polydisperse system, we need to provide information about the partial structure factors to obtain the correct dynamics \cite{voigtmann_polydis_mct}. 
Ozawa and Berthier have highlighted the fact that for a system with continuous size polydispersity the contribution from the mixing entropy term diverges \cite{ozawa}. This makes the calculation of entropy and any other dependent quantity ill-defined. They showed by calculating the inherent structure properties that when the position of the particles with similar sizes is exchanged,  the system stays in a similar basin. This modifies the vibrational entropy, which also has the same mixing entropy term. The process allowed them to group particles into a certain finite number of pseudo species leading to a finite value of the mixing entropy.
These studies thus  emphasize the importance of describing the structure of a polydisperse system in terms of the partial structure factors of the pseudo species.  

The present study attempts to develop a general framework to describe the structure of a system having continuous polydispersity. As discussed before, for a system with continuous polydispersity the number of species is the same as the number of particles which makes it difficult to describe the structure. We also know that describing all the particles in terms of a single species does not work. So the aim of  this study is to describe the system in terms of $M$ pseudo species such that the properties of this system are the same as the original system. The questions that we ask are i) Can we determine the minimum number of pseudo species $'M_{0}'$ required to describe the structure of the system? ii) Is this dependent on the property that we study? iii) Does it depend on the degree and nature of polydispersity? iv) Does it depend on the interaction potential? 

To answer these questions we use the route of calculating thermodynamic quantities which can be obtained from the structure of the liquid. Namely the potential energy of the system and the pair excess entropy. Note that the former is a linear function of the structure whereas the latter is a nonlinear function of the structure and thus can have different sensitivities to the effective structure. 
We find that by studying these above mentioned thermodynamical quantities, we can determine a value of $M_{0}$.  It depends on the type of polydispersity, the degree of polydispersity, and the interaction potential. We also provide an estimate of the width of polydispersity that can be treated like a one component system. This width appears to depend primarily on the interaction potential of the system. Systems with longer range and softer interaction potential have a better tolerance towards polydispersity. In these cases, systems with a wider spread of size can be addressed in terms of a one component system.

The organization of the rest of the paper is the following. Section II contains the simulation details.  In Section III, we discuss the methods used for evaluating the various quantities of interest. Section IV contains results with discussions and the paper ends with a brief conclusion in Section V.

\section{Simulation Details}
\begin{figure}[h]
  \centering
    \subfigure{\includegraphics[width=.45\textwidth]{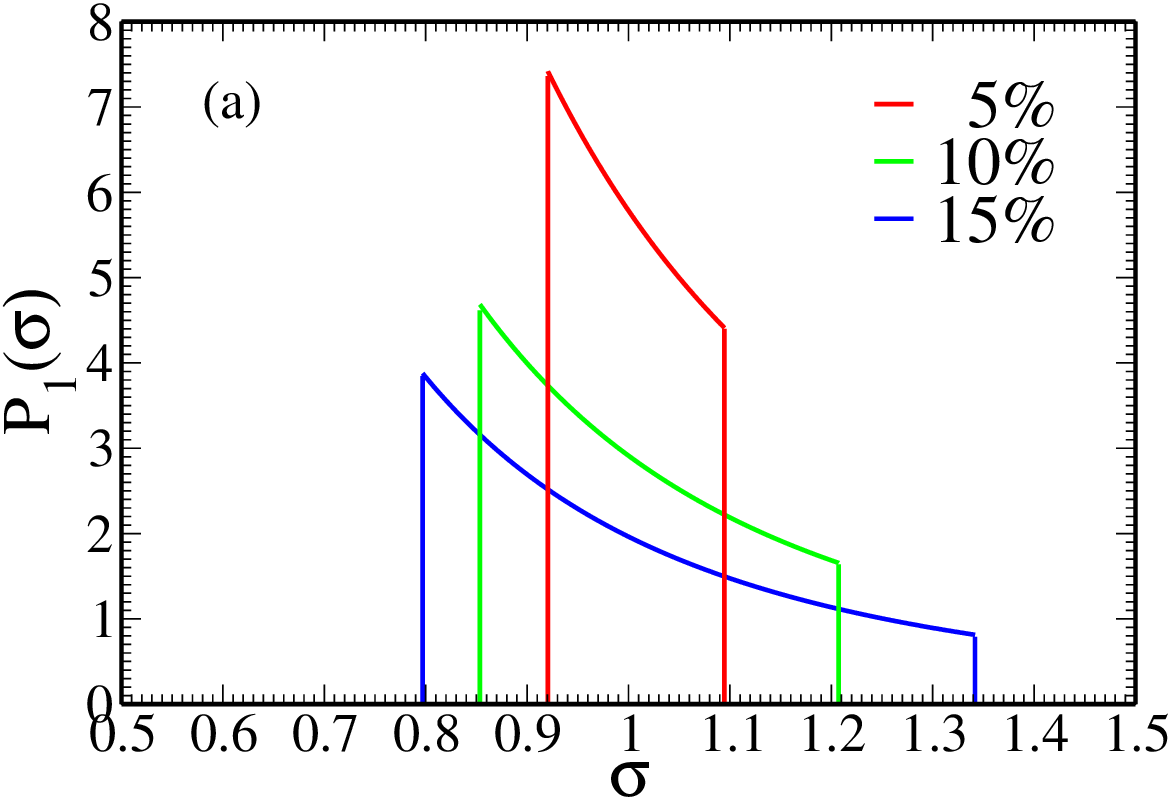}}
    \subfigure{\includegraphics[width=.45\textwidth]{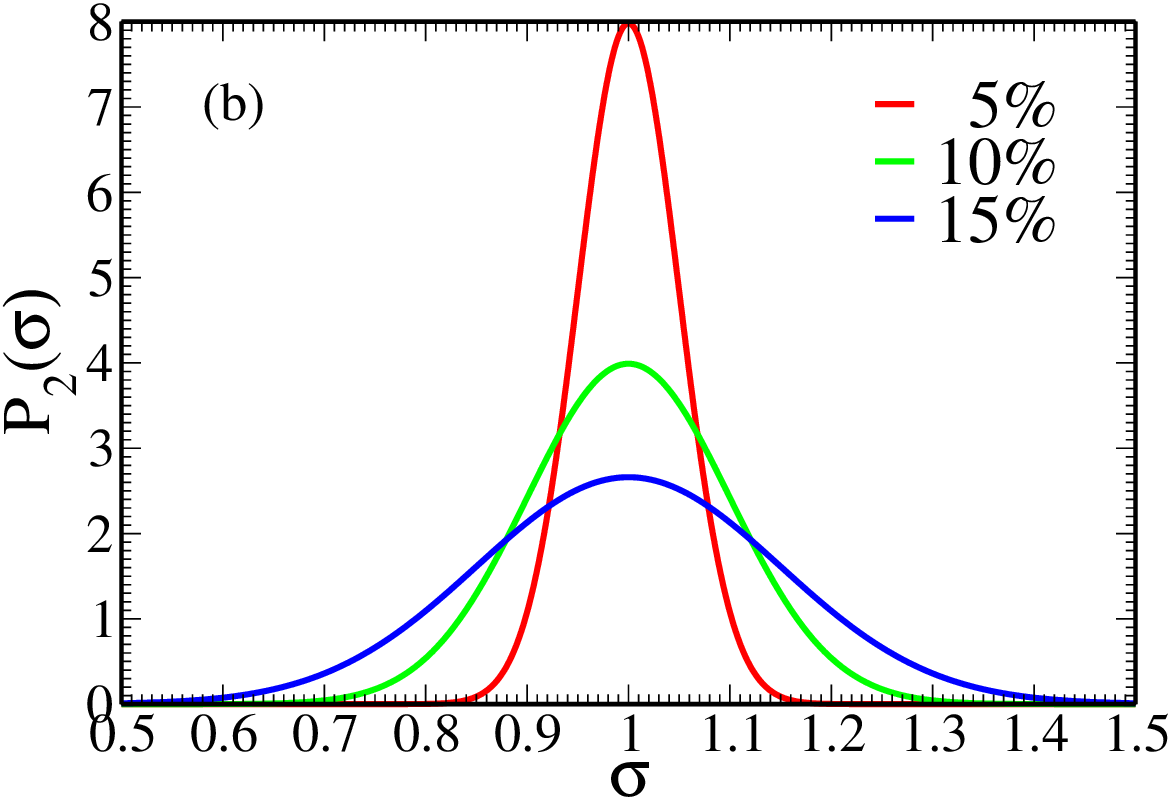}}
    \caption{\emph{Different kind of distributions (a)Constant volume fraction distribution, $P_{1}(\sigma)$ (b) Gaussian distribution, $P_{2}(\sigma)$. For the same degree of polydispersity, compared to $P_{1}(\sigma)$ the distribution is wider for $P_{2}(\sigma)$.}}
    \label{sigma_distribution}
\end{figure}

In  this  study, we perform molecular dynamics simulations for three-dimensional polydisperse system with continuous size polydispersity in the canonical ensemble. The system contains N=1000-4000 particles in a cubic box of volume V. The number density for all the systems is $\rho=N/V=1.0$. In our simulations, we have used periodic boundary conditions and Nos\'{e}-Hoover thermostat  with integration timestep 0.001$\tau$. The time constants for the  Nos\'{e}-Hoover thermostat  are taken to be 100  timesteps.  We have carried out the molecular dynamics simulations using the LAMMPS package \cite{lammps}. 
The study involves two different kinds of systems with respect to size polydispersity, constant volume fraction (CVF) and Gaussian (as shown in Fig. \ref{sigma_distribution}) and three different kinds of interaction potentials. The distributions of the particle size are continuous. This means each of the N particles has a different radius. The form of the constant volume fraction distribution is given by \cite{Bidisperse_simulation},
\begin{equation}
  P_{1}(\sigma)= \frac{A}{\sigma^3},  \hspace{1.2cm} \sigma \in [\sigma_{max},\sigma_{min}]
\label{constant_volume}  
\end{equation}
where A is the normalization constant and $\sigma_{max}$ and $\sigma_{min}$ are the maximum and minimum values of particle diameter. $\sigma_{max}$ and $\sigma_{min}$ values are given in Table.\ref{sigma_max_and_sigma_min}. The degree of polydispersity is quantified by \cite{Bidisperse_simulation} the normalized root mean square deviation 
\begin{center}
PDI = $\frac{\sqrt{<\sigma^2>-<\sigma>^2}}{<\sigma>}$ 
\label{vcf_pdi}
\end{center}
Where $\Big< ..\Big>$ defines the average of the particle size distribution. 

The Gaussian distribution is given by 
\begin{equation}
P_{2}(\sigma) = \frac{1}{\sqrt{2\pi\delta^2}}\exp^{\frac{-(\sigma-<\sigma>)^2}{2\delta^2}} 
\label{gaussian}
\end{equation}
Where $\delta$ is the standard deviation. In this distribution we consider $\sigma_{max/min}=<\sigma> \pm 3\delta$. The degree of polydispersity is quantified by,
\begin{center}
PDI =$\frac{\sqrt{<\sigma^2>-<\sigma>^2}}{<\sigma>}$ =$\frac{\delta}{<\sigma>}$    
\end{center}
For all the polydisperse systems the particle sizes are such chosen that $<\sigma>=\int P(\sigma)\sigma d\sigma$ = 1 and is kept as the unit of length for all the systems studied here.

\begin{table}
\caption{\emph{Details of the size distributions, constant volume fraction and Gaussian, for different degrees of polydispersity $PDI$=$\frac{\sqrt{<\sigma^2>-<\sigma>^2}}{<\sigma>}$. The maximum, $\sigma_{max}$, and minimum $\sigma_{min}$, values of the diameter of the particles. The volume fraction $\eta$ is also given showing an increase in $\eta$ with degree of polydispersity.}}
\begin{tabular}{|c|c|c|c|c|c|}
    \hline
Distribution &  PDI \% & $\sigma_{max}$ & $\sigma_{min}$ & $\Delta\sigma$ & $\eta$\\
\cline{1-6}
$P_1(\sigma)$ & 5\% & 1.1 &  0.92 & 0.18 & 0.53 \\
\cline{2-6}
& 10\% & 1.21  & 0.85  & 0.36  & 0.54 \\
\cline{2-6}
& 15\% & 1.34  & 0.8  & 0.54 & 0.56 \\
\cline{1-6}
$P_2(\sigma)$ &  5\%  & 1.15 & 0.85  & 0.3 & 0.53\\
 \cline{2-6}
 & 10\%  & 1.3 & 0.7 & 0.6 & 0.54 \\
\cline{2-6}
& 15\%  & 1.45  & 0.55 & 0.9 & 0.56  \\
\cline{1-6}
\end{tabular}

\label{sigma_max_and_sigma_min}
\end{table}

The three different interaction potentials studied here are inverse power law (IPL) potential, Lennard-Jones  (LJ) potential and its repulsive counterpart the Weeks-Chandeler-Andersen (WCA) potential.
The inverse power law potential (IPL) between two particles i and j  is given by, \cite{Bidisperse_simulation}$^,$\cite{tridisperse_simulation}.

\begin{equation}
 U(r_{ij}) =
 \begin{cases}
 \epsilon_{ij} (\frac{\sigma_{ij}} {r_{ij}})^{12} + \sum_{l=0}^{2} c_{2l} (\frac{r_{ij}} {\sigma_{ij}})^{2l}, & (\frac{r_{ij}} {\sigma_{ij}}) \leq x_{c}\\
 0, & (\frac{r_{ij}} {\sigma_{ij}}) > x_{c}
\end{cases} 
\label{potential} 
\end{equation}
The constants $c_0 , c_2$ and $c_4 $ are selected such that the potential becomes continuous up to its second derivative at the cutoff $x_c=1.25\sigma_{ij}$. 

The LJ potential between the two particles i and j  is described using truncated and shifted LJ potential\cite{LJ_potential};

\small\begin{equation}
 U(r_{ij})=
\begin{cases}
 U^{(LJ)}(r_{ij};\sigma_{ij},\epsilon_{ij})- U^{(LJ)}(r^{(c)}_{ij};\sigma_{ij},\epsilon_{ij}),    & r_{ij}\leq r^{(c)}_{ij}\\
   0,                                                                                       & r_{ij}> r^{(c)}_{ij}
\end{cases}
\label{kalj_eq}
\end{equation}
where $U^{(LJ)}(r_{ij};\sigma_{ij},\epsilon_{ij})=4\epsilon_{ij}[(\frac{\sigma_{ij}}{r_{ij}})^{12}-(\frac{\sigma_{ij}}{r_{ij}})^{6}]$. The cutoff for the LJ system is $r^{(c)}_{ij}=2.5\sigma_{ij}$ and for the WCA system is $r^{(c)}_{ij}=2^{1/6}\sigma_{ij}$.\cite{WCA_potential}
The interaction strength between two particles i and j is  $\epsilon_{ij}$ = 1.0. $\sigma_{ij}=\frac{(\sigma_i+\sigma_j)}{2}$, where  $\sigma_{i}$ is a diameter of particle i and it varies according to the system. Length, temperature and times are given in units of $<\sigma>, \epsilon_{ij}$ and $\big(\frac{m<\sigma>^2}{\epsilon_{ij}}\big)^\frac{1}{2}$ respectively. For all state points, the equlibration is performed for $100 \tau_{\alpha}$ ($\tau_{\alpha}$ is the $\alpha$- relaxation time) and three to five independent samples are analyzed. 
As discussed later, in this work both potential energy and pair excess entropy are calculated using the partial structure factors. In a system with continuous polydispersity, each particle has a different diameter. So there are $\frac{N\times N}{2}$ partial radial distribution functions. For $N=1000$ there are thus $500,000$ partial rdf and with increase in $N$ this number grows as $N^{2}$. Calculating these many partial rdf with good precision is an impossible task. However, in this study, we divide the total system into $M$ species. Particles with diameter range $(\sigma_{max}-\sigma_{min})/M$ are treated as a single species. Note that this is an approximation because the particle sizes in a single species are still different. With an increase in $M$ this diameter range becomes narrow and the approximation leads to less error. The maximum value of $M$ that we have used in this work is 26. Thus we have calculated at the most $338$ partial rdf.  Although the study is performed in the high temperature regime where the production run length is usually around  100 $\tau_{\alpha}$ where $\tau_{\alpha}$ varies between $5-100$, for this study to get a good precision of the partial rdf we require longer production run lengths to compensate for the poor particle averaging. For 5\% PDI the production run length is $10^7$ and for 15\% PDI  and 10\% PDI the production run lengths are $10^7$ for T= 5.0-0.38 and $6\times 10^{7}$ for T=0.36-0.2

\section{Definition and Background}

\subsection{Radial distribution function}

The partial radial distribution  $g_{\alpha\beta}(r)$ is define as:
\begin{equation}
g_{\alpha\beta}(r) =\frac{V}{N_{\alpha}N_{\beta}}\Big<\sum_{i=1}^{N_{\alpha}} \sum_{j=1, j \neq i}^{N_{\beta}} \delta (r- r_i +r_j) \Big>  
 \label{gr_2}
\end{equation}
\noindent
Where V is the volume of the system, $N_{\alpha}$ , $N_{\beta}$ are the number of $\alpha$ and $\beta$ type of particles, respectively.

The effective one component radial distribution function, g(r) can be written in terms of partial rdf of $M$ species as.\cite{gr_hansen_mcdonald}
\begin{equation}
 g(r) = \sum_{\alpha,\beta=1}^{M} \chi_{\alpha} \chi_{\beta} g_{\alpha\beta}(r)
 \label{gr}
\end{equation}

Where $\chi_\alpha$ and $\chi_\beta$ are the mole fraction of $\alpha$ and $\beta$ particles, respectively.

\subsection{Potential energy}
The per particle potential energy of the system can be exactly calculated from simulation $E_{sim}$. 
The same can also be written in terms of the partial radial distribution function, $E_{2}$.

\begin{equation}
E_{2} = \frac{\rho}{2}\sum_{\alpha,\beta=1}^{M} \chi_{\alpha} \chi_{\beta}\int_{0}^{\infty} 4\pi r^2 U_{\alpha\beta}(r) g_{\alpha\beta}(r) dr   
\label{E_2}
\end{equation}

In the effective one component treatment the energy can be written as $E_{2}^{eff}$,
\begin{equation}
E_{2}^{eff} = \frac{\rho}{2}\int_{0}^{\infty} 4\pi r^2 U(r) g(r)dr 
\label{E_2_eff}
\end{equation}

\subsection{Excess entropy}
Excess entropy $S_{ex}$ is a loss of entropy due to the interaction between the particle or in other words excess entropy is a difference between $S_{total}$ and $S_{ideal}$ at same temperature T and density $\rho$. The value of $S_{ex}$ is always a negative. $S_{ex}$ can be evaluated by thermodynamic integration (temperature density landscape)\cite{exc_entropy}. Entropy at high temperature and low density is that of an ideal gas entropy. This $S_{ideal}$ is a relative reference for any other state points entropy calculation. Other state point entropy can be calculated using a combination of isotherms (Eq.\ref{exc_1}) and isochoric (Eq.\ref{exc_2}) paths, making sure that no phase transitions occurs along the selected path\cite{exc_entropy}.\\

\begin{equation}
 S_{ex}(T,V') - S_{ex}(T,V) = \frac{U(T,V') - U(T,V)}{T} + \int_{V}^{V'} \frac{P(V)}{T} dV
 \label{exc_1} 
 \end{equation}

\begin{equation}
S_{ex}(T',V') - S_{ex}(T,V') = \int_{T}^{T'} \frac{1}{T} \big(\frac{\delta U}{\delta T}\big)_{V'} dT
\label{exc_2}     
\end{equation}
   
Addition of Eq. \ref{exc_1} and Eq. \ref{exc_2} give the total excess entropy.

\subsection{Pair Entropy}
Using Kirkwood factorization\cite{Kirkwood}, the excess entropy can also be expressed in terms of an infinite series,
\begin{equation}
S_{ex} = S_2 + S_3 + S_4 .... 
\label{s2}
\end{equation}
\noindent
Where $S_n$ is an entropy contribution due to n particle spatial correlation. The pair excess entropy, $S_2$ includes 80\% of the total excess entropy\cite{S_2_and_S_exc_are_equal}. We can calculate $S_2$ from the partial rdf of $M$ species \cite{exc_entropy}-

\begin{equation}
\small\frac{S_2}{k_B}=-2\pi \rho \sum_{\alpha,\beta = 1}^{M} \chi_{\alpha} \chi_{\beta} \int_{0}^{\infty}r^2dr\{g_{\alpha\beta}(r)\ln g_{\alpha\beta}(r) - (g_{\alpha\beta}(r) - 1)\}
\label{S_2}  
\end{equation}

Where $k_B$ is a Boltzmann constant.

If we do not consider the different species then the entropy for an effective one component system can be written as $S_{2}^{eff}$,

\begin{equation}
\frac{S_{2}^{eff}}{k_B}=-2\pi \rho \int_{0}^{\infty}r^2dr\{g(r)\ln g(r) - (g(r) - 1)\}
\label{S_2_eff}  
\end{equation}

\subsection{Onset temperature calculation from Inherent structure energy }
While cooling a glass forming liquid from high temperatures at the onset temperature, $T_{onset}$ the system's thermodynamic and dynamic properties deviate from its high-temperature behavior. There are different dynamical and thermodynamical measures of $T_{onset}$. The temperature predicted by each method is not identical but lies in a similar range. A comparison of the different methods is given in Ref.\onlinecite{onset_crosspoint}. In this work, we will discuss the one calculated from the inherent structure energy and the other from the excess entropy.
 
The inherent structure energy is the potential energy evaluated at the local minimum of the energy reached from the configuration via the steepest descent procedure. As suggested earlier \cite{inherent_sastry} the onset temperature is connected to the inherent structure energy. At high temperatures as the system is not influenced by the landscape properties, the average inherent structure energy is almost temperature independent. However below a certain temperature, where the landscape properties influence the system, the inherent structure energy starts to decrease rapidly. Usually, the two different regimes are fitted to two straight lines and the point where these lines cross is identified as the onset temperature, $T_{onset}$. 

\subsection{Dynamics}
 In this work to characterize the dynamics we consider the self part of the overlap function $Q(t)$ definied as,\cite{overlap_shiladitya}

\begin{equation}
Q(t) =\frac{1}{N}\sum_{i=1}^{N} \langle \omega (|{\bf{r}}_i(t)-{\bf{r}}_i(0)|)\rangle \quad ,
\label{eq22}
\end{equation}

\noindent
where the function $\omega(x)$ is 1 if $0\leq x\leq a$ and $\omega(x)=0$ otherwise. The parameter $a$ is chosen to be 0.3, a value that is slightly larger than the size of the cage. 

Note that the dynamics can also be obtained from self intermediate scattering function $F_s(q,t)$ where $q=2\pi/r_{max}$, $r_{max}$ being the position of the first peak of the radial distribution function. Since relaxation times from $Q(t)$ and $F_s(q,t)$ behave very similarly at low temperature, we use $Q(t)$ for the dynamics. 

\section{Result}

\subsection{Effective one-component description}

 As discussed before it is a common practice to describe the structure of a polydisperse system in terms of an effective one component system.

\begin{figure}[h]
    \centering
    \includegraphics[width=.45\textwidth]{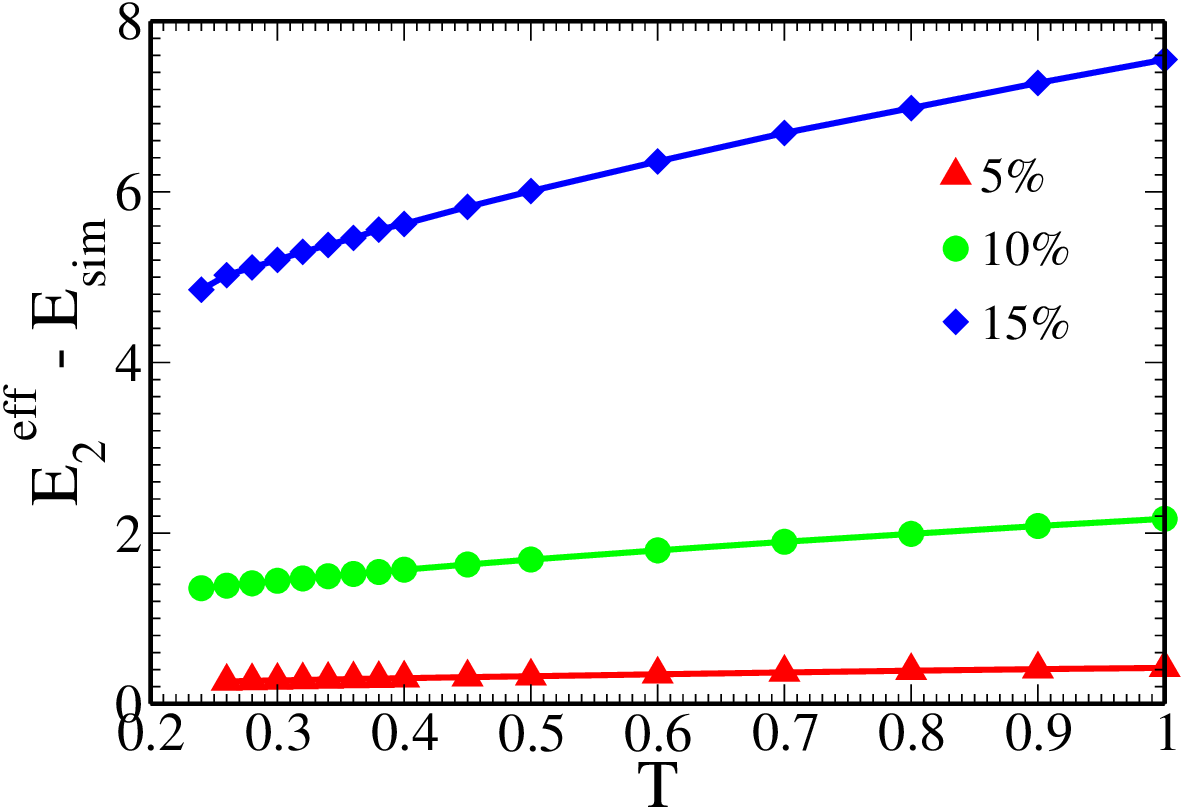}
    \caption{\emph{The difference in energy obtained from effective one component radial distribution functions, $E^{eff}_{2}$ and simulation,$E_{sim}$ as a function of T at different PDIs.} Here particles are interacting via IPL potential and the particle size distribution is given by $P_{1}(\sigma)$ (constant volume fraction distribution). }
    \label{all_system_E}
\end{figure}

\begin{figure}[h]
    \centering
    \includegraphics[width=.45\textwidth]{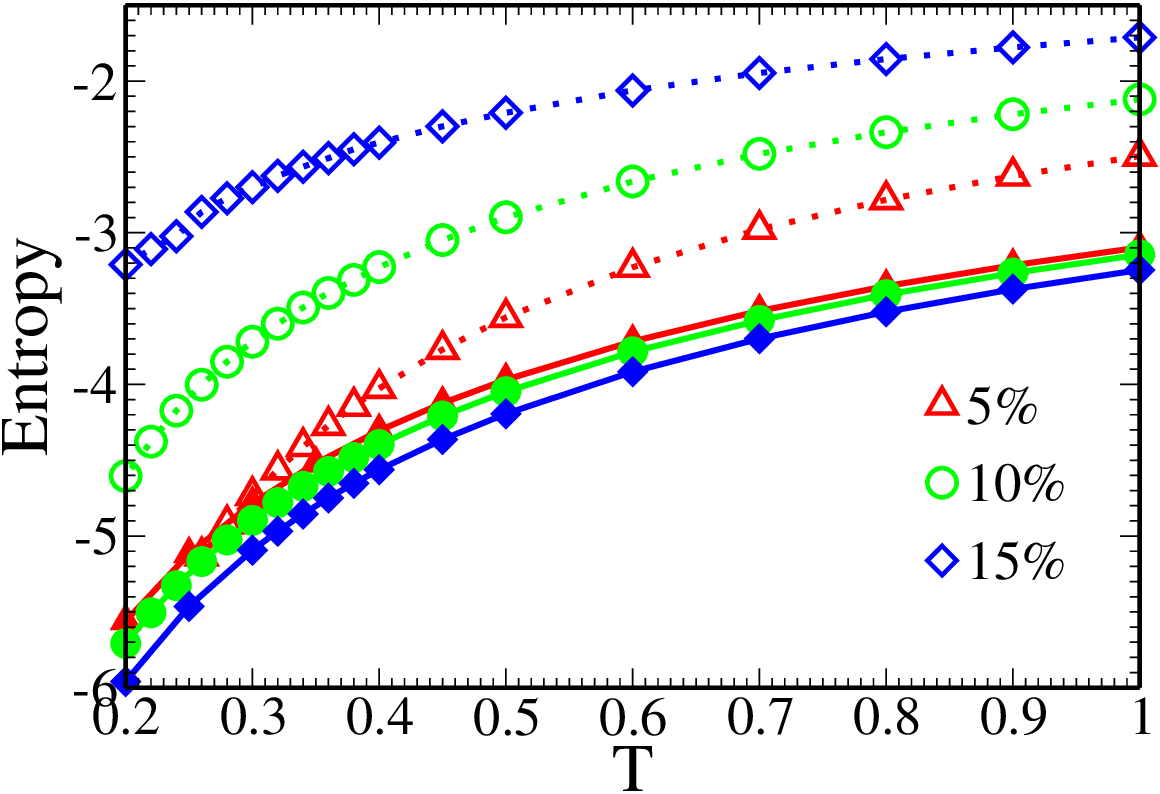}
    \caption{\emph{Excess entropy, $S_{ex}$ and effective one component pair excess entropy, $S^{eff}_2$ (Eq.\ref{S_2_eff}) at different PDIs. Solid line with filled symbol represents $S_{ex}$ and dotted line with open symbol represents  $S^{eff}_{2}$. Here particles are interacting via IPL potential and the particle size distribution is given by $P_{1}(\sigma)$ (constant volume fraction distribution).   } }
    \label{all_S_exc}
\end{figure}

In Fig.\ref{all_system_E}, we plot the difference between the average per particle potential energy of the species agnostic $E^{eff}_{2}$ and that obtained from simulation $E_{sim}$ for systems with different PDI values (5$\%$,10$\%$ and 15$\%$). In the simulation study, the particle sizes are obtained from $P_{1}(\sigma)$ distribution and they interact via IPL potential. In Fig.\ref{all_S_exc} we also plot the $S_{ex}$ and the species agnostic $S^{eff}_{2}$ for the above mentioned systems.
Note that if the structure is described properly then $E_{sim}=E_{2}$
and $S_{2}$ is not exactly equal to $S_{ex}$ but comprises of $80\%$ of its value \cite{onset_crosspoint,S_2_and_S_exc_are_equal,s2_ne_s_exc_1st,s2_ne_s_exc_2nd,s2_ne_s_exc_3rd,s2_ne_sexc_4th,s2_ne_s_exc_5th,exc_entropy}.  

\begin{figure}[h]
    \centering
   \subfigure{\includegraphics[width=.4\textwidth]{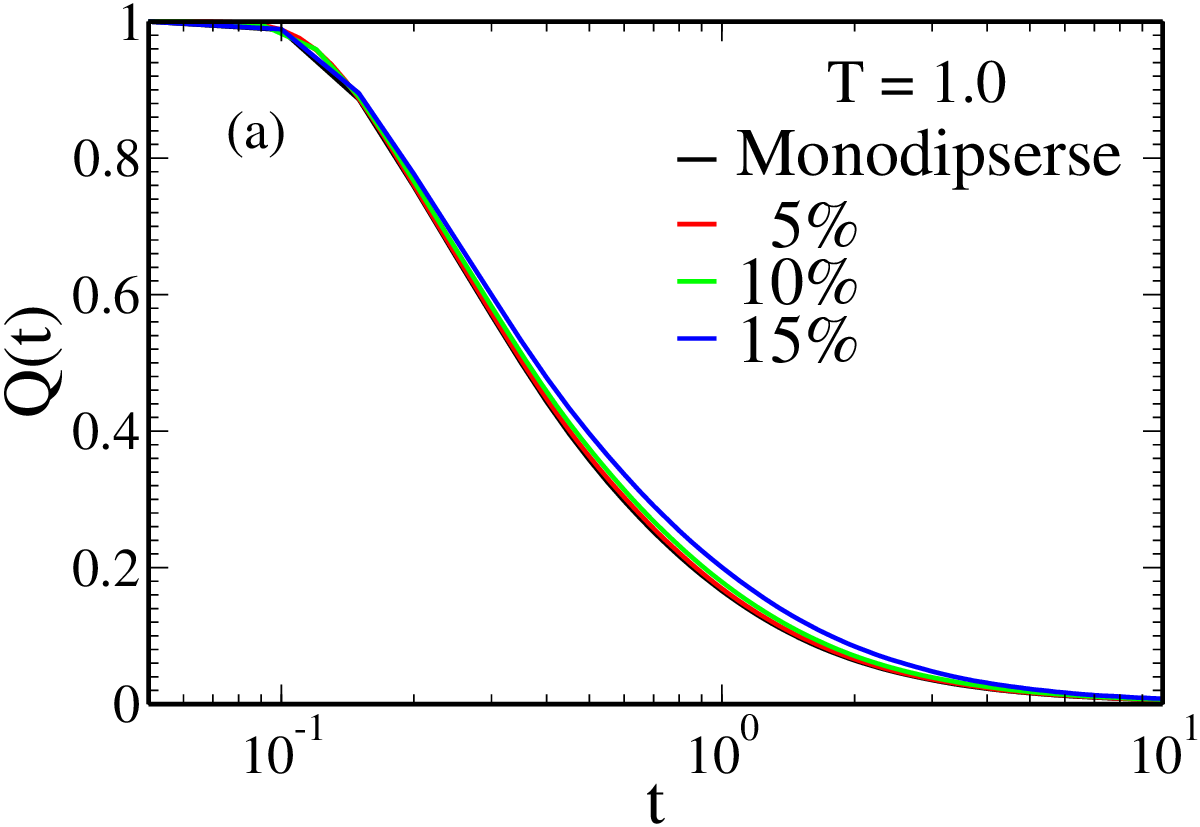}
    \label{dynamic_same_volume}}
    \subfigure{\includegraphics[width=.4\textwidth]{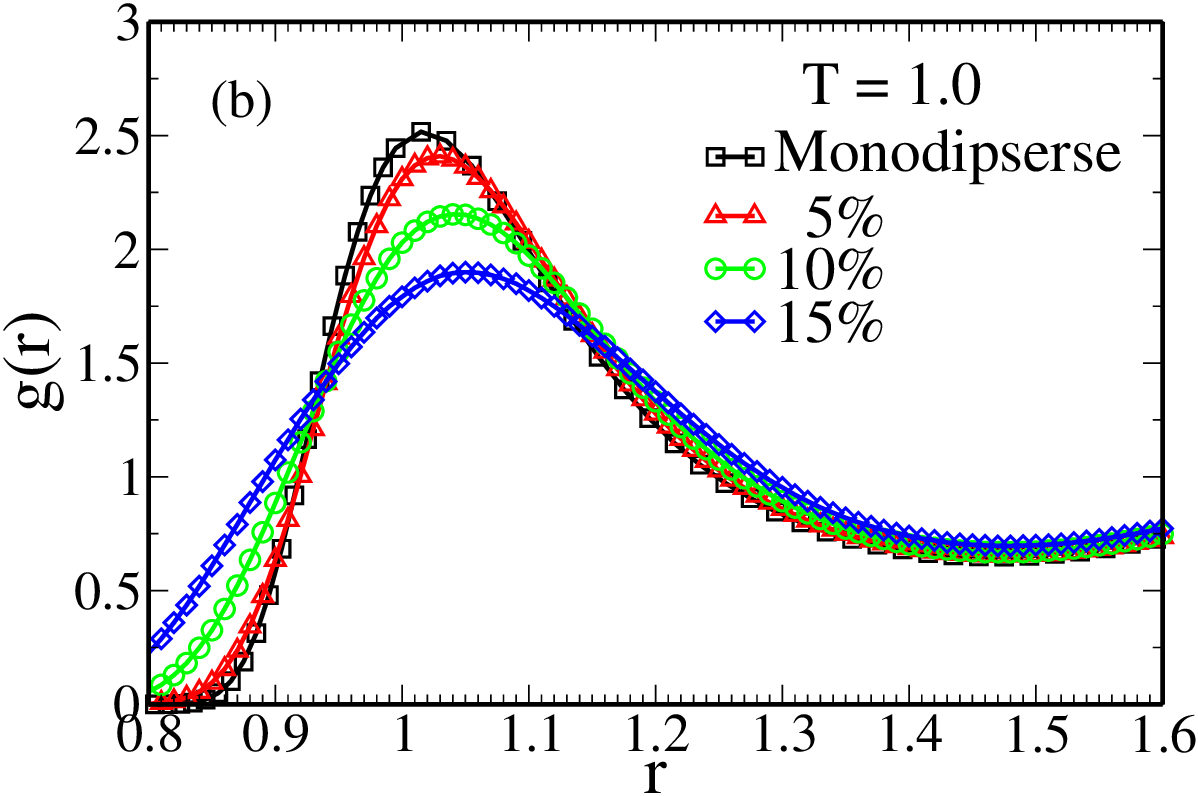}
    \label{softness_gr_T_1.0}}
     \subfigure{\includegraphics[width=.4\textwidth]{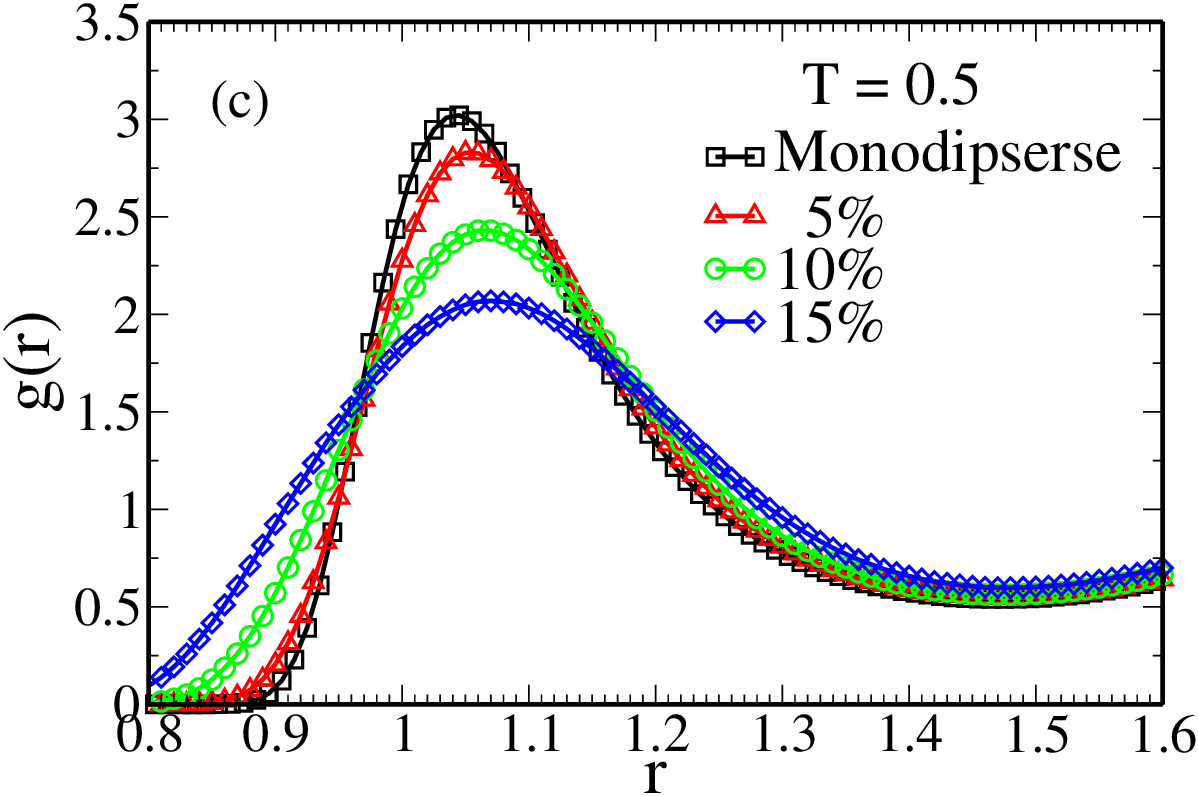}
    \label{softness_gr_T_0.5}}
        \caption{\emph{(a) Dynamics of systems at different PDIs. The overlap function is plotted against time. (b) Effective one component radial distribution function of the systems at T = 1.0.  (c) Same as Fig.\ref{softness_gr_T_1.0} at T = 0.5}. Black square,red triangle, green circle and blue diamond represent a mono disperse system, 5\% PDI, 10\% PDI and 15\% PDI, respectively. With an increase in PDI although the dynamics remains almost the same the structure shows a substantial softening. Here particles are interacting via IPL potential and the particle size distribution is given by $P_{1}(\sigma)$ (constant volume fraction distribution).}
    \label{softness}
\end{figure}

We find that as the PDI increases the difference between $E_{sim}$ and $E^{eff}_{2}$ and $S_{ex}$ and $S^{eff}_{2}$ increases. This clearly shows that as expected, with an increase in PDI the effective one component description of the system becomes less accurate. In Fig.\ref{softness} we plot both the dynamics and the effective one component rdf of the systems. We find that within the temperature range studied here although the dynamics remains almost the same, with an increase in polydispersity the structure appears to soften.We have plotted the rdf at two temperatures, (T=1.0 and 0.5) and it appears that the softening is present in both temperatures. However, the fact that the difference between $E^{eff}_{2}$ and $E_{sim}$ reduces at low temperatures (Fig.\ref{all_system_E}) do suggest that the softening also reduces with temperature. This artificial softening of the structure leads to the increase in $E^{eff}_{2}$ and $S^{eff}_{2}$. Note that even for $5\%$ polydispersity we find that the effective one component structure of the system fails to provide the correct value of the potential energy and the pair excess entropy. These results presented in Fig.\ref{all_system_E} and Fig.\ref{all_S_exc} are not surprising but a confirmation of the observations made earlier \cite{main_paper_Truskett,voigtmann_polydis_mct}. 

\subsection{Pseudo species and its dependence on degree of polydispersity}

Describing the structure of a continuous polydisperse system can be challenging?
Unlike in a discrete polydisperse system where each species has a finite number of particles and all of them have the same size, for a continuous polydisperse system the number of species is the same as the number of particles. However, let us assume that we describe a pseudo system where we divide the particles into M number of pseudo species (where $M<N$) in terms of the size of the particles. 
In doing so we bunch particles with similar but different sizes, in a group and assign an average size to them. 
This introduces disparity in the actual size and the assigned size of the particles and leads to an error in describing the properties of the system. An extreme case of that ($M=1$) can be seen in Fig.\ref{all_system_E} and Fig.\ref{all_S_exc}.
For a fixed $M$, the maximum difference in the actual diameter of a particle and its assigned average diameter is $\Delta\sigma/2M$.
Thus with an increase in M this error reduces and at $M=N$ the pseudo system is exact. So the first question is can we describe the structure of a system in terms of an optimum number of species $M_{0}$, where $M_{0}<<N$ such that the structure can provide a correct estimate of the thermodynamic quantities? If we can then how does $M_{0}$ depend on the degree of polydispersity?

\begin{figure}[h]
    \centering
    \subfigure{\includegraphics[width=.45\textwidth]{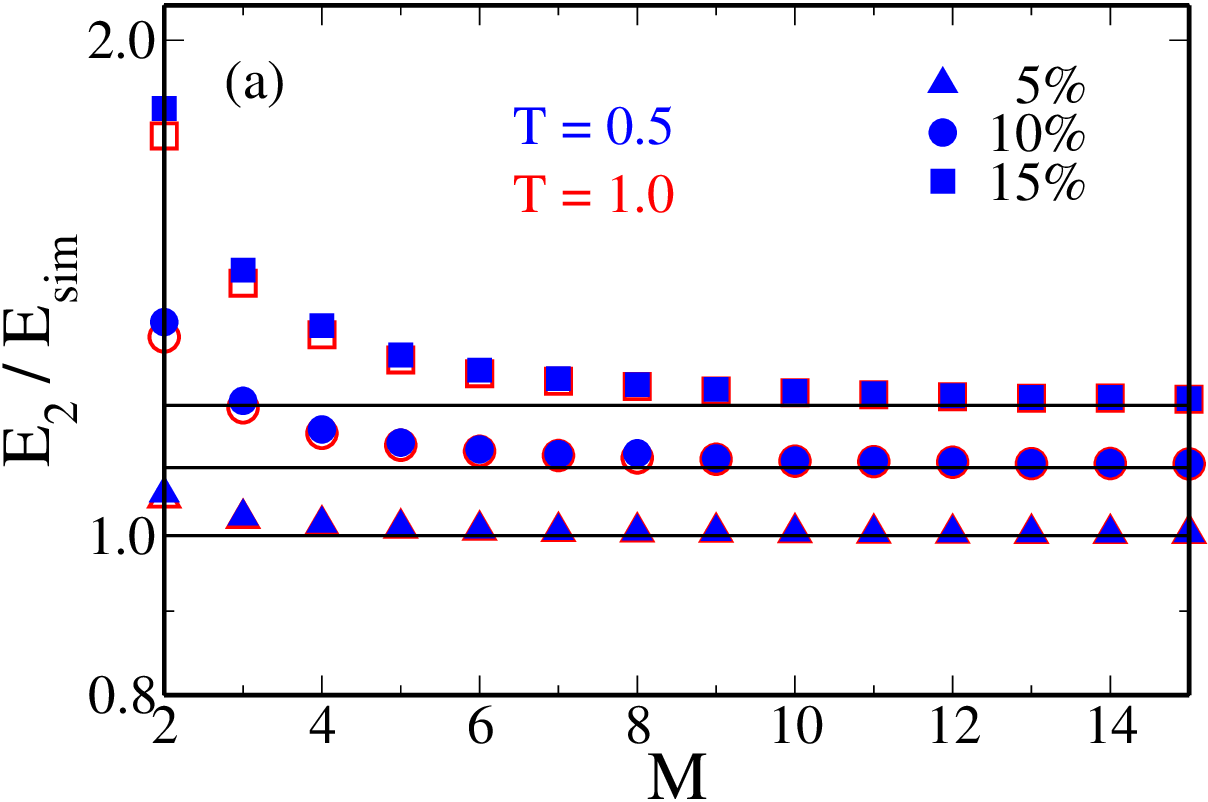}}
      \subfigure{\includegraphics[width=.45\textwidth]{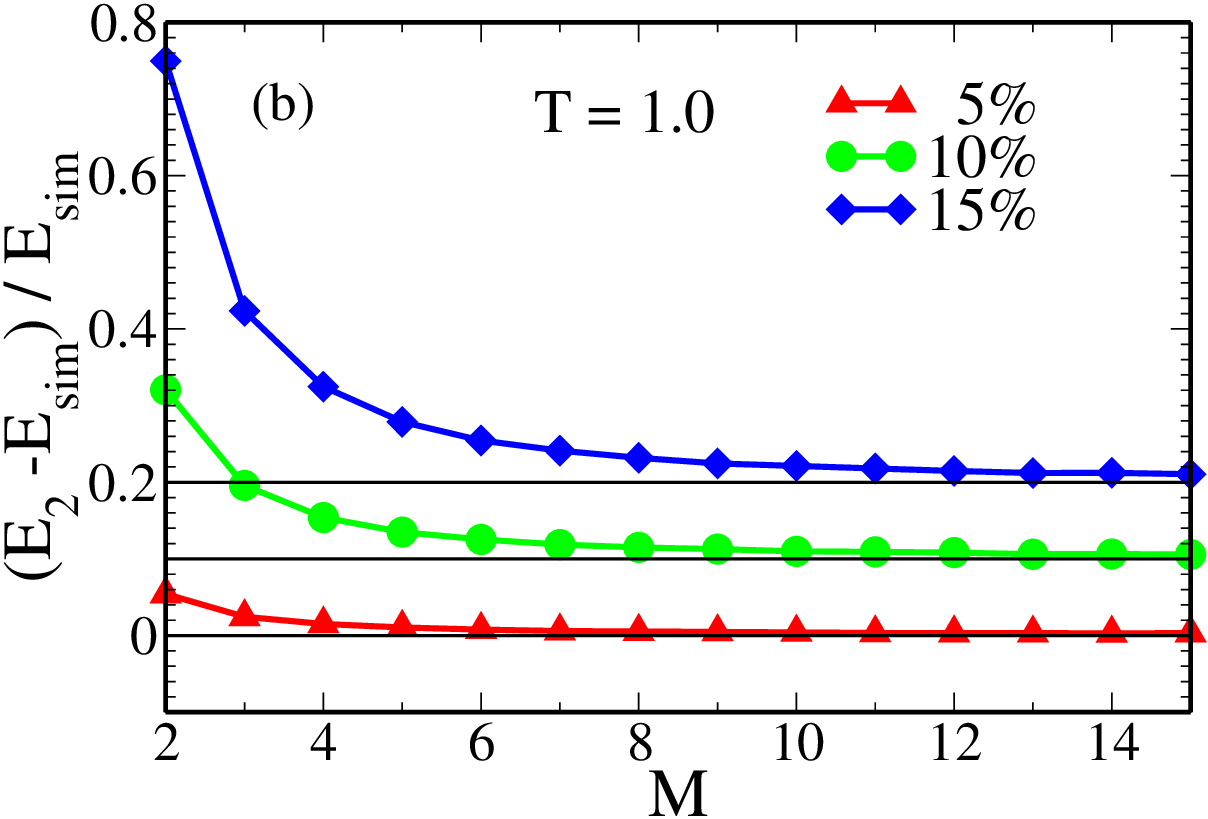}}
    \caption{\emph{Comparison between energy obtain from simulation, $E_{sim}$, and energy obtain from partial radial distribution functions, $E_{2}$ (Eq. \ref{E_2}). (a)Ratio of $E_{2}$ and $E_{sim}$ vs the number of pseudo species $M$ at T = 1.0 (open red symbols) and T = 0.5 (filled blue symbols) for different PDIs.(b) Relative error calculation between $E_{sim}$ and $E_{2}$, ($E_{2}-E_{sim})/E_{sim}$ plotted as a function of $M$ for different PDIs. 
   For better visualization, we have shifted the y-axis of the 10\% PDI plot by 0.1, and 15\% PDI plot by 0.2. The horizontal lines signify the corresponding large M values which are 1.0 for (a) and 0.0 for (b).} Here particles are interacting via IPL potential and the particle size distribution is given by $P_{1}(\sigma)$ (constant volume fraction distribution).}
   \label{E2byEsim}
\end{figure}

In Fig.\ref{E2byEsim}(a) we plot $\frac{E_{2}}{E_{Sim}}$ as a function of $M$, at two different temperatures for the different PDIs. For systems with a fixed value of PDI as we increase the value of M the $E_{2}$ decreases and after a certain value of M, $E_{2} \simeq E_{sim}$. We find that this is weakly temperature dependent. 
For this work, we consider that at $T=1$ the minimum number of pseudo species for which $\frac{(E_{2}-E_{sim})}{E_{sim}} <0.01$ is $M_{0}$.  The value of $M_{0}$ is system dependent and as expected increases with the increase in PDI value as can be seen from Fig.\ref{E2byEsim}(b). Note that while determining $M_{0}$ this choice of the relative error (0.01) is arbitrary but practical. In principle, we can choose values much smaller or probably larger than this. However, later, while discussing the value of $M_{0}$ as obtained from entropy, we will see that this choice is reasonable.

\begin{figure}[h]
    \centering
    \subfigure{\includegraphics[width=.45\textwidth]{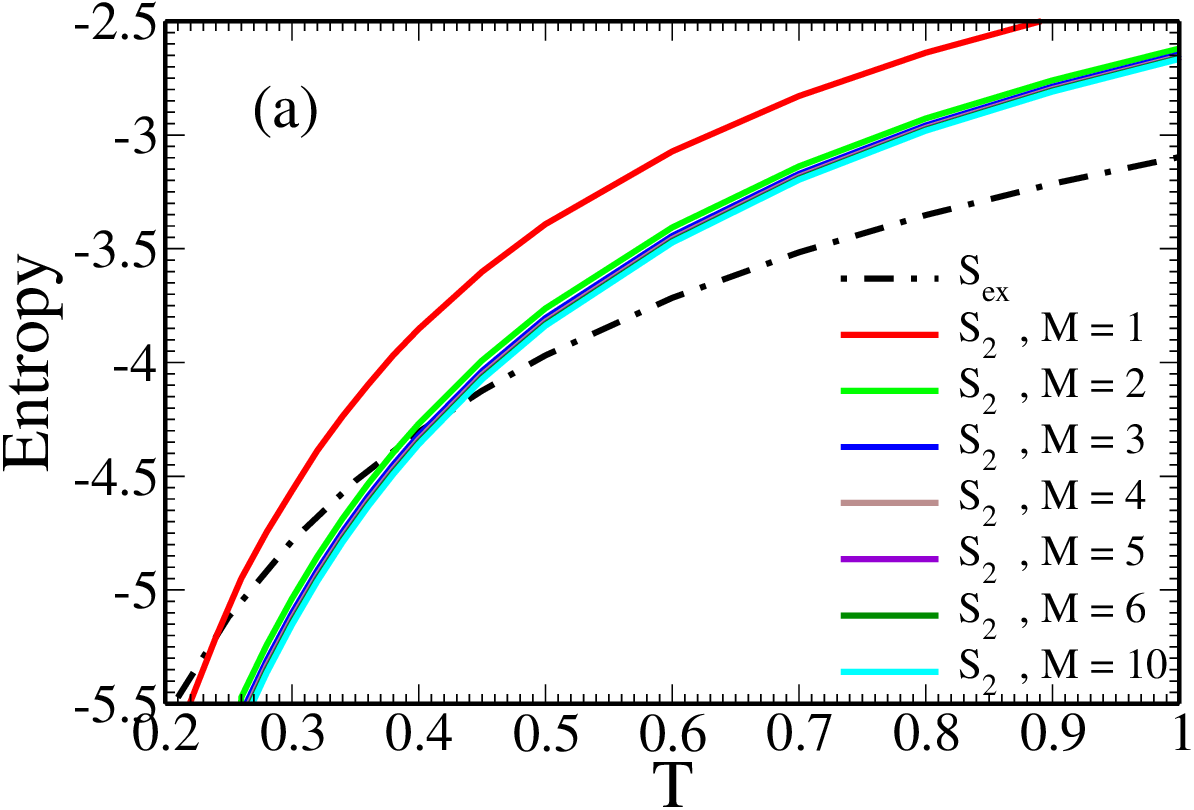}}
    \subfigure{\includegraphics[width=.45\textwidth]{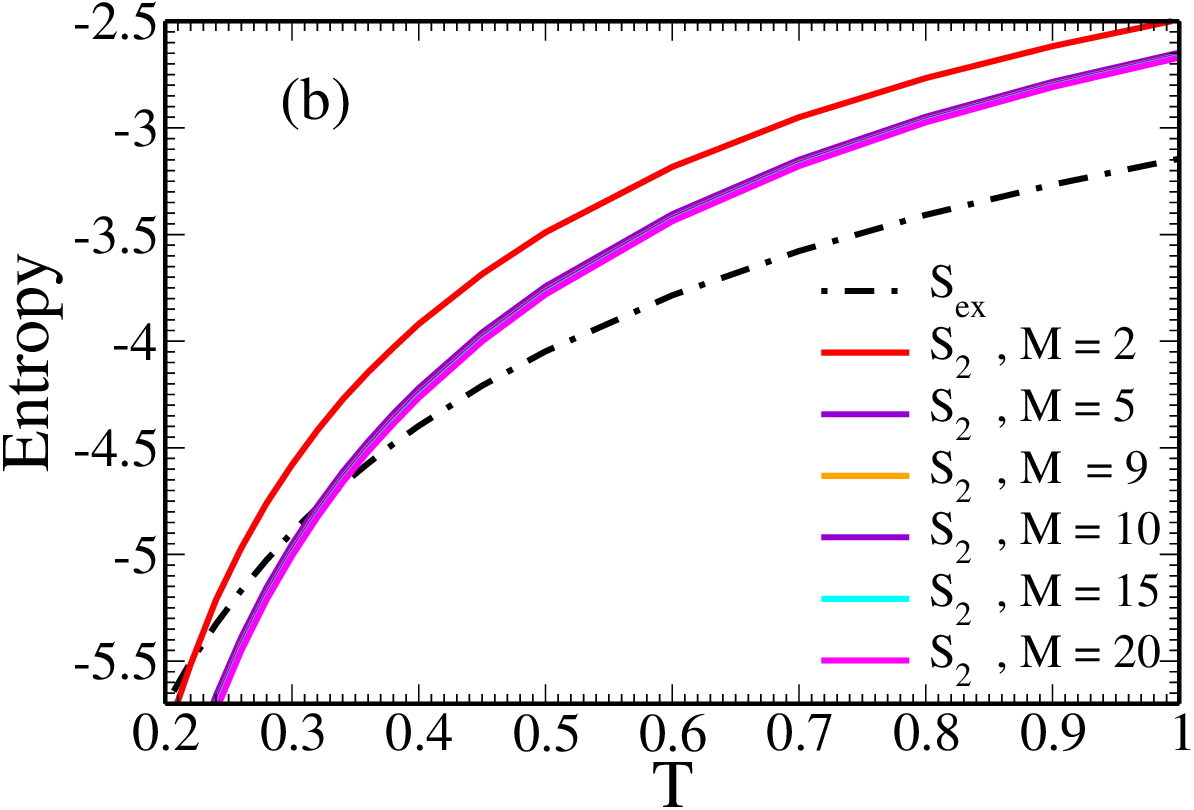}}
   \subfigure{\includegraphics[width=.45\textwidth]{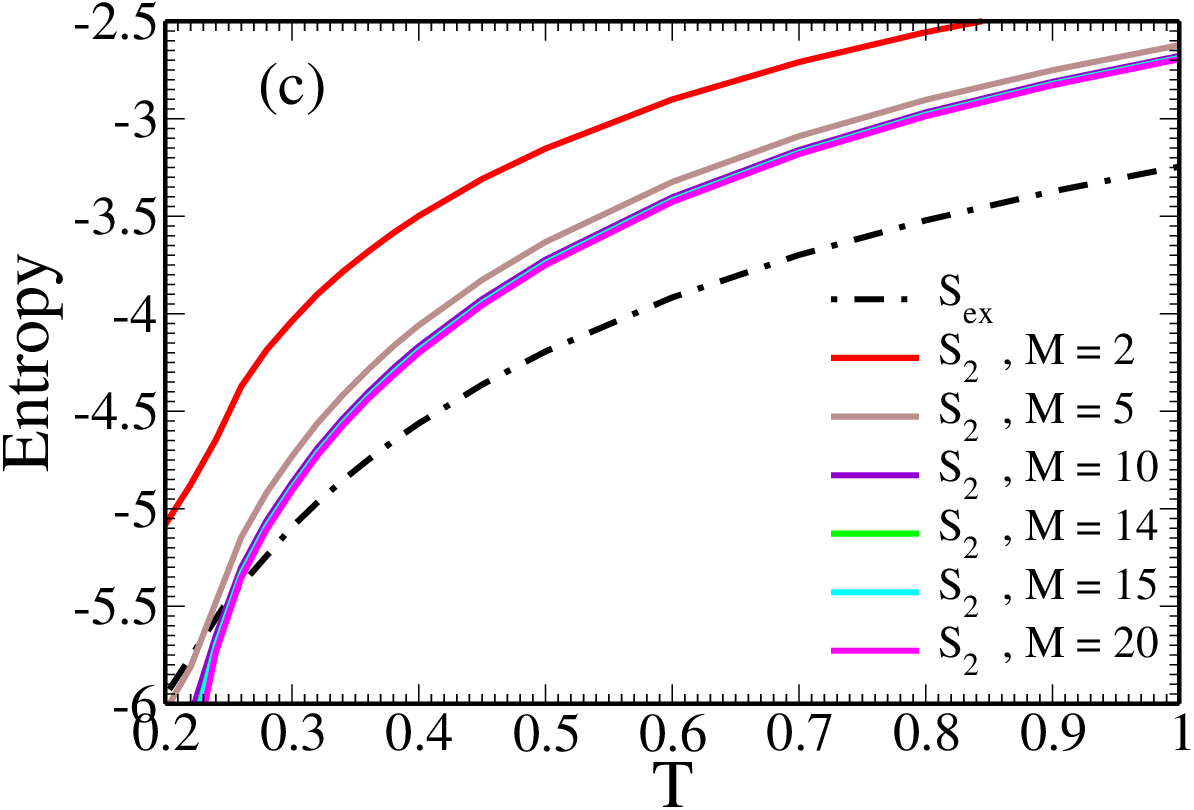}}
    \caption{\emph{Excess entropy, $S_{ex}$, and pair excess entropy $S_2$. The latter is calculated at different values of $M$ (Eq.\ref{S_2}). Dashed dot line represents $S_{ex}$ and solid lines represent $S_{2}$.  (a) PDI 5\% (b) PDI 10\% (c) PDI 15\%. Here particles are interacting via IPL potential and the particle size distribution is given by $P_{1}(\sigma)$ (constant volume fraction distribution). }}
 \label{crosspoint_PDI}
\end{figure}

\begin{figure}[h]
    \centering
    {\includegraphics[width=.45\textwidth]{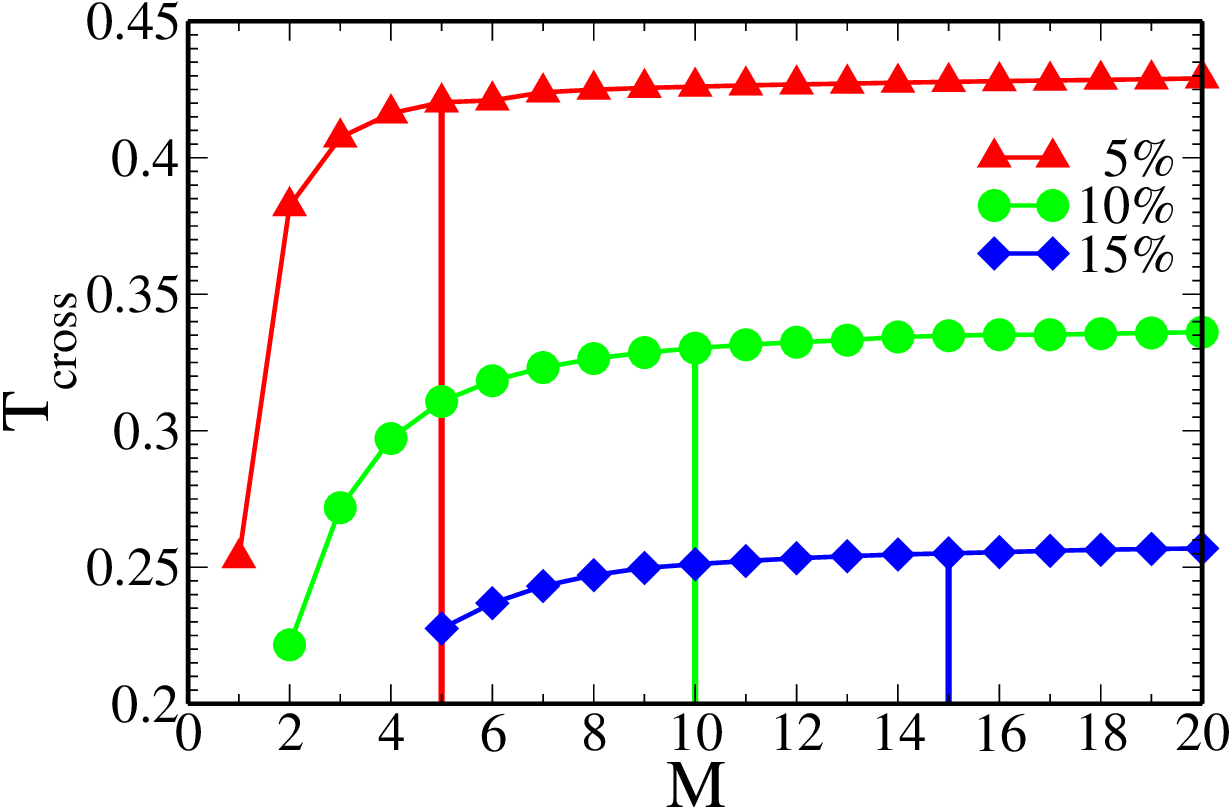}}
    \caption{\emph{$T_{cross}$ vs $M$ plot for different PDIs.  Initially, $T_{cross}$ increases with M but after certain value of M, it saturates. For larger PDI the saturation takes place at a higher $M$ value. The vertical lines give the value of $M_{0}$ obtained from energy criteria. Here particles are interacting via IPL potential and the particle size distribution is given by $P_{1}(\sigma)$ (constant volume fraction distribution).}}
    \label{M_vs_T_cross_Cv}
\end{figure}

Is the value of $M_{0}$ sensitive to the thermodynamic quantity that we calculate or is it universal? To answer this question we calculate the two body pair entropy for different values of $M$. We find that similar to the energy, as $M$ increases the $S_{2}$ comes closer to $S_{ex}$ (Fig.\ref{crosspoint_PDI}). However, even for large values of $M$, $S_{2}$ is not equal to $S_{ex}$. This is because unlike the potential energy which can be exactly calculated in terms of the partial rdf, only a part of the excess entropy can be calculated from the rdf (Eq.\ref{s2})  \cite{onset_crosspoint,S_2_and_S_exc_are_equal,s2_ne_s_exc_1st,s2_ne_s_exc_2nd,s2_ne_s_exc_3rd,s2_ne_sexc_4th,s2_ne_s_exc_5th,exc_entropy}. This makes it difficult to use the same methodology as used for potential energy to make an estimation of $M_{0}$ from entropy.

However, from our earlier studies, we know that if the structure of the liquid is described properly then the excess entropy and the two body pair entropy crosses each other at a temperature,$T_{cross}$ which can be considered as the onset temperature of the supercooled liquid dynamics \cite{onset_crosspoint}. This onset temperature can also be obtained from the change in the slope of the temperature dependence of the inherent structure energy \cite{inherent_sastry} and also other methods \cite{onset_crosspoint}. As shown earlier the values of the onset temperatures obtained using these different methods are not exactly the same but they are in a  similar range \cite{onset_crosspoint}.     

In Fig.\ref{M_vs_T_cross_Cv} we plot the variation of $T_{cross}$ with $M$ for the different systems. For higher PDI,  at small values of $M$, we cannot calculate $T_{cross}$ which implies that $S_{2}$ is far away from $S_{ex}$ and never crosses it as seen in Fig.\ref{crosspoint_PDI}. However, from our other estimates of onset temperature, we know that we are in a temperature range where these two forms of entropy should cross. As $M$ increases the two functions cross at some temperature $T_{cross}$. We find that initially $T_{cross}$ increases with $M$ and then after a certain value of $M$ it shows a saturation. As mentioned before $S_{2}$ is not the total excess entropy of the system. There is no other method of calculating $S_{2}$. Thus it is not possible to do a similar error estimation of pair excess entropy as done for the potential energy. However, the saturation of $T_{cross}$ is an indication of the saturation of $S_{2}$ w. r. t. $M$.  We find that this saturation value of $T_{cross}$ is in a similar range as the estimated onset temperature using the method of inherent structure energy (see Sec III E and Table \ref{table_2}). In this plot we also mark the $M_{0}$ values as obtained from the potential energy. We find that the $M$ value for which $T_{cross}$ saturates falls in the similar range as $M_{0}$. The values of $T_{cross}$ at $M=M_{0}$ and the $T_{onset}$ are given in Table \ref{table_2}. Thus we show that the minimum number of pseudo species required to describe the potential energy of the system can also describe the two body excess entropy of the system. 
Note that although with $M_{0}$ pseudo species we can get a reasonable value of $S_{2}$, this quantity is not the total excess entropy of the system. The residual multi-particle entropy (RMPE) defined as the difference between the total excess entropy and the pair excess entropy, $S_{ex}-S_{2}$ although has a small value when compared to $S_{2}$, plays an important role in describing the thermodynamics of the system. For example, it has been observed that if we ignore RMPE then the correlation between dynamics and thermodynamics expressed via the well known Adam-Gibbs relation does not hold \cite{unraveling}. It has also been observed that in the supercooled liquid regime RMPE provides us a measure of the activated dynamics of the system \cite{unraveling,gcm_manoj}. Thus although the pseudo species description provides us a reasonable estimation of $S_{2}$, care should be taken while using this quantity in describing the full thermodynamics of the system.

The details of the $M_{0}$ values for the different systems are given in Table \ref{table_2}. We also tabulate a quantity $\Delta\sigma_{0}=\frac{\Delta \sigma}{M_{0}}$. We find that although $M_{0}$ is dependent on the PDI this quantity $\Delta \sigma_{0}$ is not. Note that when $M=M_{0}$, the maximum error in assigning a diameter to a particle is $\frac{\Delta \sigma_{0}}{2}$. Thus our study suggests that the thermodynamic properties of the system studied here are not sensitive to a change in diameter by $\frac{\Delta \sigma_{0}}{2}$ and for a constant volume fraction polydisperse system interacting via IPL potential when $\Delta \sigma \approx 0.036$ we can treat it as a monodisperse system.

\begin{figure}[h]
    \centering
    \subfigure{\includegraphics[width=.45\textwidth]{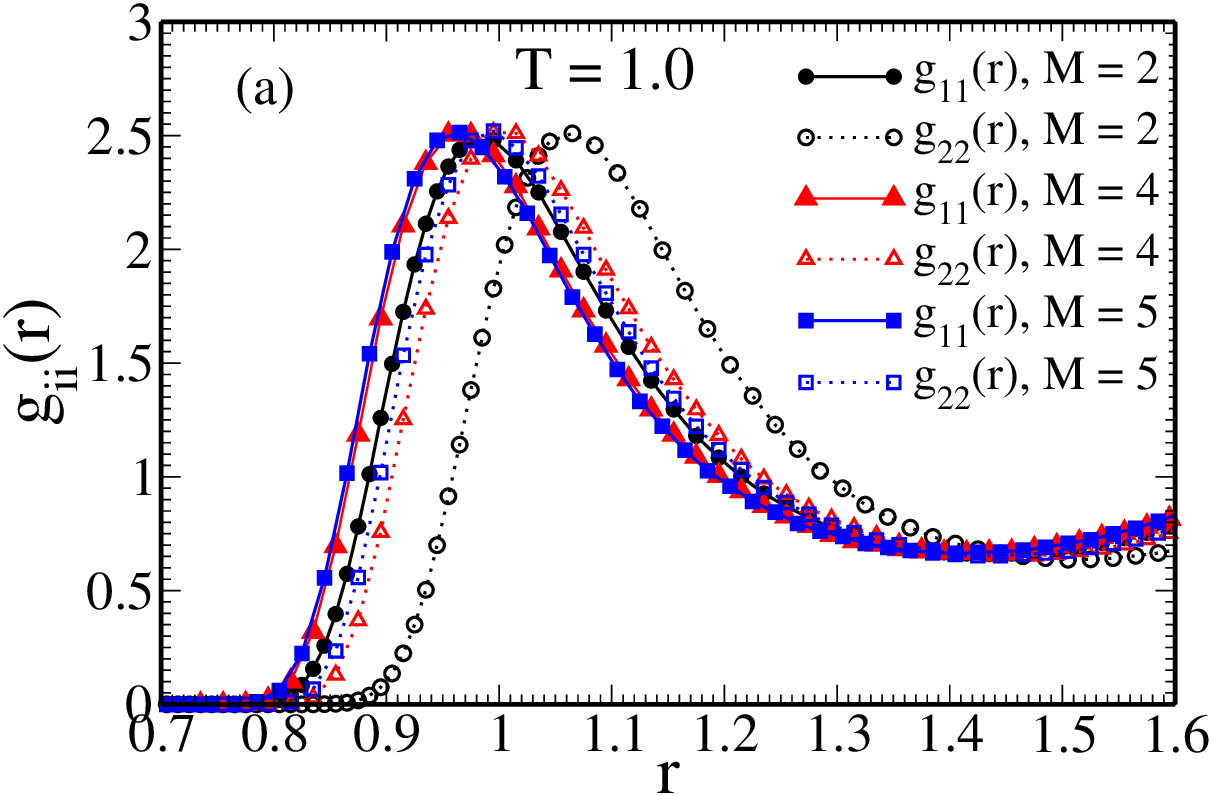}}
    \subfigure{\includegraphics[width=.45\textwidth]{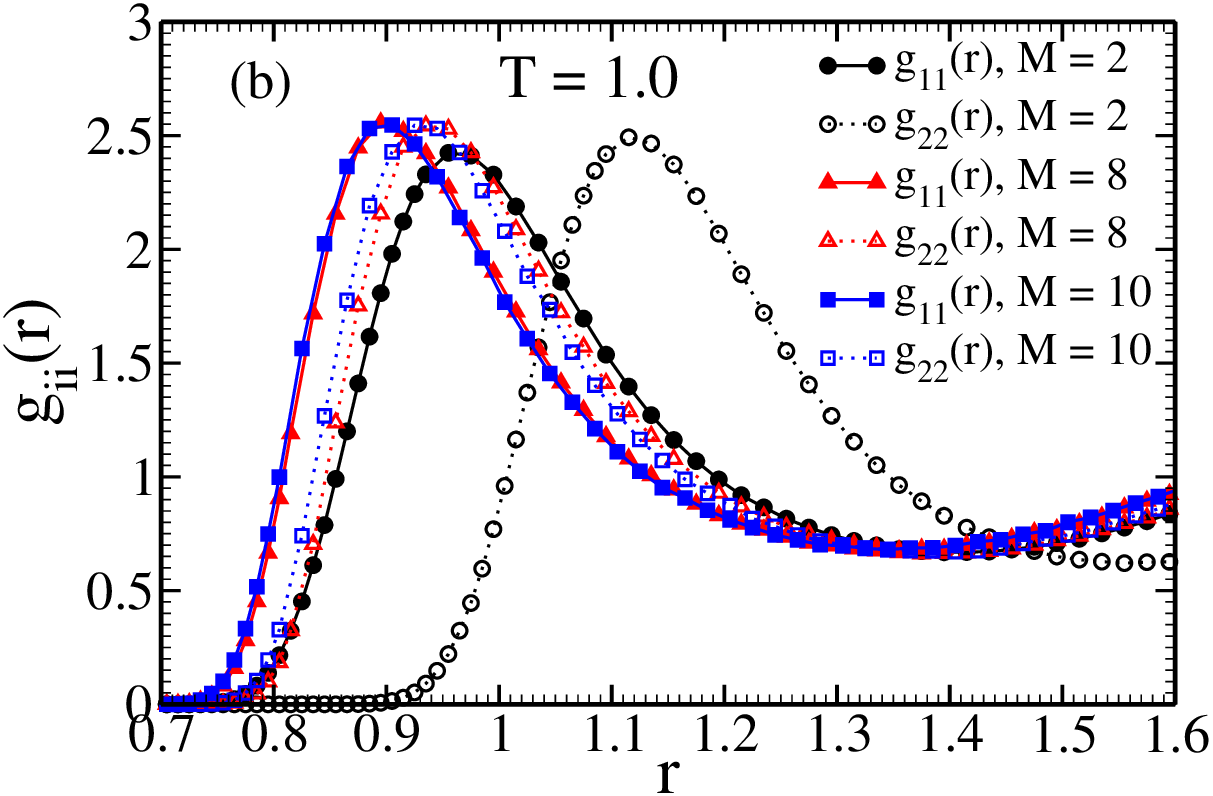}}
    \subfigure{\includegraphics[width=.45\textwidth]{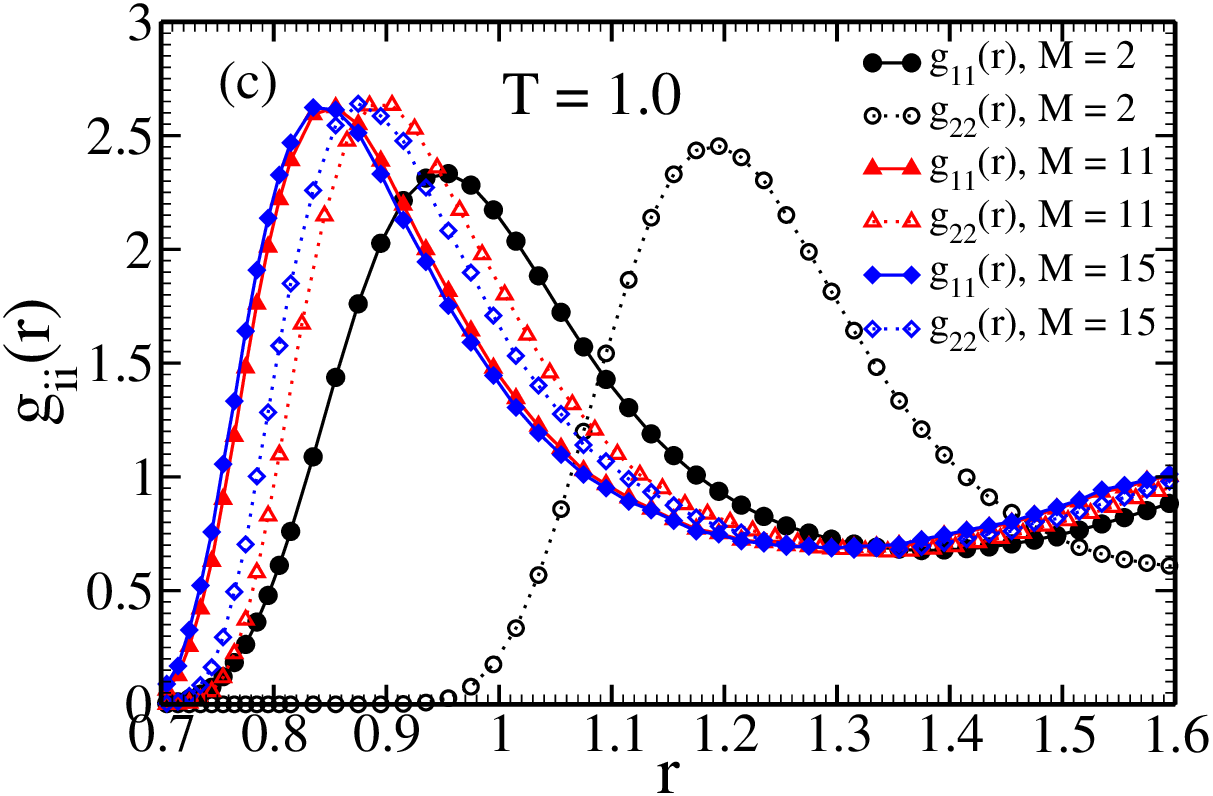}}
       \label{gr_data}
       \caption{\emph{The partial radial distribution function for the first two species for different values of $M$. (a)PDI 5\% (b) PDI 10\% (c) PDI 15\%. For $M=M_{0}$ the rdf peak of the two consecutive species almost overlap. Here particles are interacting via IPL potential and the particle size distribution is given by $P_{1}(\sigma)$ (constant volume fraction distribution).}}
\label{gr_PDI}       
\end{figure}

Interestingly we find that when we plot the partial rdf for two consecutive pseudo species, $g_{11}(r)$ and $g_{22}(r)$ (here these two species 1 and 2 have the largest and the second largest number of particles, respectively) for different values of $M$ then for $M=M_{0}$ the peaks of the two rdfs almost overlap (Fig.\ref{gr_PDI}) . Thus we can say that when the size difference of the two consecutive species are such that there is a large overlap between the radial distribution functions of two consecutive species then they can be treated as a single species.

\begin{table}[h]
\begin{center}
\addtolength{\tabcolsep}{-1pt}
\caption{\emph{The values of $M_{0}$ and $\Delta\sigma_{0}$ for different systems. We also provide the values of $T_{cross}$ at $M=M_{0}$ and for comparison we give the $T_{onset}$ values obtained from fitting the Inherent structure energy to two straight lines.}}
\begin{tabular}{|c|c|c|c|c|c|c|}
\hline
Distribution & Potential &  PDI $\%$  & $M_{0}$ & $\Delta\sigma_{0}$=$\frac{\Delta\sigma}{M_{0}}$ & $T_{cross}(M_{0})$ & $T_{onset}$\\ 
\cline{1-7}
$P_{1}(\sigma)$ & IPL
&  5\%  & 5   & 0.036 & 0.42 &0.43  \\
\cline{3-7}
& & 10\%  & 10 & 0.036 & 0.33 & 0.36 \\
\cline{3-7}
& & 15\%  & 15  & 0.036 &0.26 & 0.31\\
\cline{1-7}
$P_{2}(\sigma)$ & IPL
 & 5\%  & 7 & 0.043 & 0.43 & 0.46\\
 \cline{3-7}
 & & 10\%  & 14  & 0.043 & 0.35 & 0.34\\
\cline{3-7}
& & 15\%   & 21  & 0.043 & 0.28 & 0.30 \\
\cline{1-7}
 $P_{1}(\sigma)$ & WCA & 15\% & 20 & 0.027 & 0.58 & 0.7 \\ 
\cline{2-7}
 & LJ  & 15\% & 12 & 0.045 & 0.67 & 0.81 \\
 \hline
\end{tabular}

\label{table_2}

\end{center}
\end{table}

\subsection{Effect of the type of distribution on $M_{0}$ and $\Delta\sigma_{0}$ }

\begin{figure}[h]
    \centering
    \includegraphics[width=.45\textwidth]{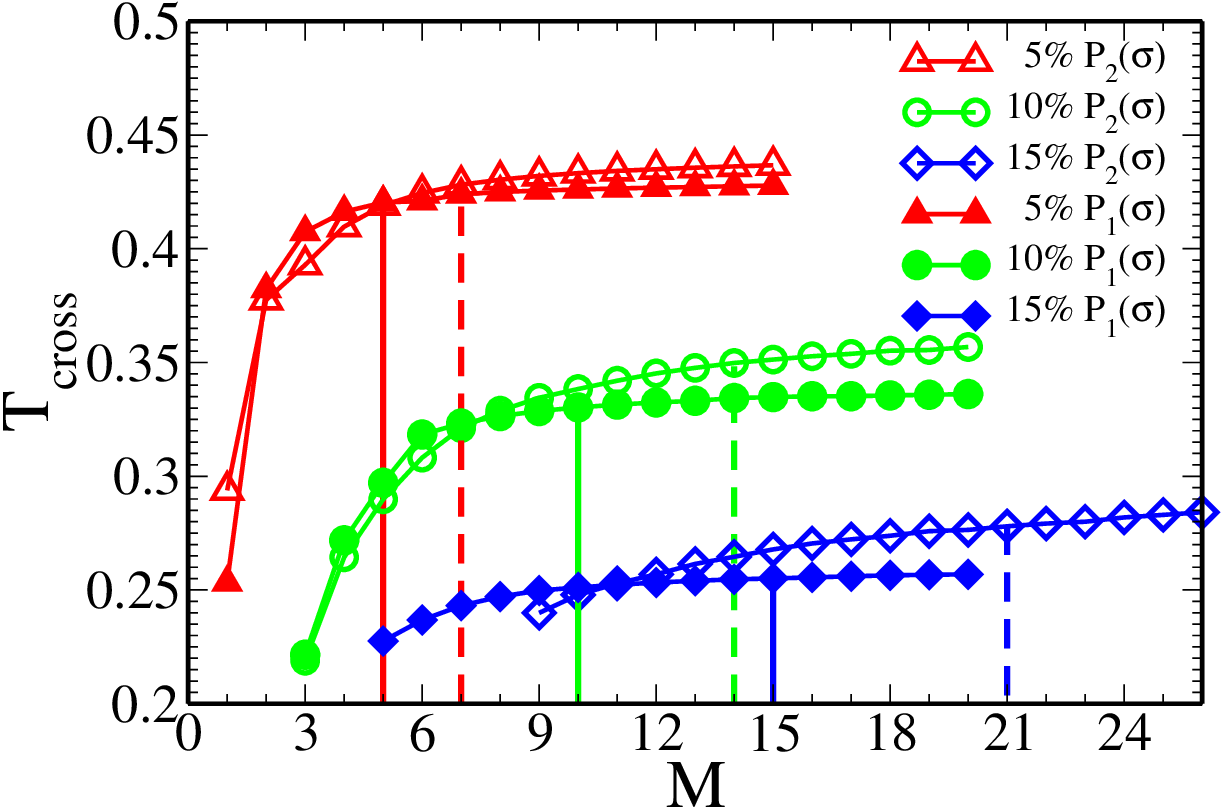}
    \caption{\emph{The effect of the type of distribution on $M_0$. $T_{cross}$ vs $M$ plot for different PDIs for systems where particles are interacting via IPL potential and the particle size distribution are given by $P_{1}(\sigma)$ (constant volume fraction distribution) and $P_{2}(\sigma)$ (Gaussian distribution). The vertical lines give the value of $M_{0}$ obtained from energy criteria (solid lines for $P_{1}$ and dashed lines for $P_{2}$ distribution). At the same value of PDI for the Gaussian distribution the $M_{0}$ is higher and $T_{cross}$ saturates at a higher $M$ value.}}
    \label{All_data}
\end{figure}

We next study the effect of the type of distribution on $M_{0}$ and $\Delta\sigma_{0}$. 
In Fig.\ref{All_data} we compare the $M$ dependence of $T_{cross}$ for systems where polydispersity is described in terms of $P_{1}(\sigma)$ (constant volume fraction) and $P_{2}(\sigma)$ (Gaussian), for three different PDIs. In the same plot the $M_{0}$ values as obtained from the potential energy are also marked. 

At the same PDI, the nature of saturation of $T_{cross}$ and also the value of $M_{0}$ are different for the two different distributions. Compared to the constant volume fraction distribution, the values of $M_{0}$ are higher for the Gaussian distribution. 
The reason behind this can be understood by comparing Fig.\ref{constant_volume} and Fig.\ref{gaussian} (also see Table \ref{sigma_max_and_sigma_min}). For the same PDI, the Gaussian distribution is wider compared to the constant volume fraction distribution. A closer observation tells us that the saturation of $T_{cross}$ is better for the CVF distribution when compared to the Gaussian distribution. Note that for the Gaussian distribution $M_{0}$ is higher (more number of species), and by nature towards the tail of the distribution the number of particles is less so the partial rdf for the largest and the smallest species are poorly averaged. We have seen that with an increase in system size the saturation improves (not shown here).

We find that when compared to CVF distribution, the $\Delta \sigma_{0}$ is marginally higher for the systems with Gaussian distribution (see Table \ref{table_2}).
Note that we divide a continuous polydisperse system into $M$ species in a way that the difference in diameter of two consecutive species is always separated by $\frac{\Delta\sigma}{M}$. This implies that the percentage error in calculating the size of the smaller particles are higher than that of the larger particles. Also, note that in constant volume fraction distribution the smallest particles are the largest in number thus by construction the error is maximum for the dominant species. On the other hand for the Gaussian distribution, the place where the percentage error is maximum the population of particles are the minimum. Thus compared to the Gaussian distribution for continuous volume fraction we need to go to marginally smaller values of $\Delta\sigma_{0}$. A way to increase $\Delta \sigma_{0}$ (decrease $M_{0}$) in CVF distribution can be to have a size dependent bin width such that the percentage error in describing the size of a smaller particle is the same as that of a larger particle.

\subsection{Effect of interaction potential on $M_{0}$}
\begin{figure}
    \centering
    \includegraphics[width=.45\textwidth]{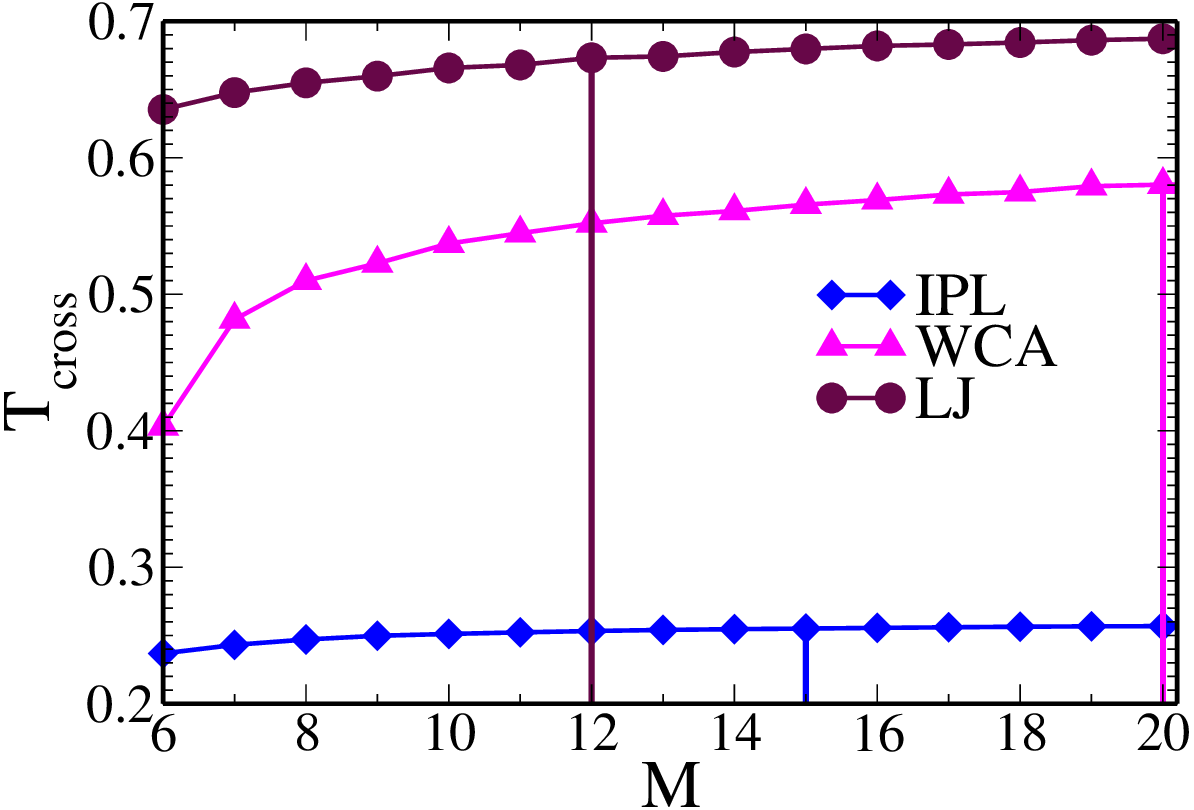}
    \caption{\emph{Role of interaction potential on $M_0$. $T_{cross}$ vs $M$ plot for 15$\%$ PDIs for systems where particles are interacting via IPL, WCA and LJ potential and the particle size distribution are given by $P_{1}(\sigma)$ (constant volume fraction distribution). The vertical lines give the value of $M_{0}$ obtained from energy criteria. The value of $M_{0}$ is higher for WCA potential and the $T_{cross}$ also saturates at higher value of $M$.}}
    \label{WCA_IPL_LJ_onset}
\end{figure}

Next, we study the role of interaction potential on the value of $M_{0}$ (also $\Delta\sigma_{0}$) and the  saturation of $T_{cross}$. 
For this we study $P_{1}(\sigma)$ system, with PDI=15$\%$ and vary the interaction potential between the particles (IPL, WCA, and LJ). 
The parameter values are given in Table \ref{table_2}. When we compare the IPL, WCA and LJ systems we find that $M_{0}$ value is higher for the WCA system and also the $T_{cross}$ of the WCA system appears to saturate at a higher $M$ value (see Fig.\ref{WCA_IPL_LJ_onset}). This suggests that $\Delta\sigma_{0}$ for the WCA potential is smaller than the LJ and the IPL systems (see Table.\ref{table_2}). 

\begin{figure}

    \centering
    \includegraphics[width=.45\textwidth]{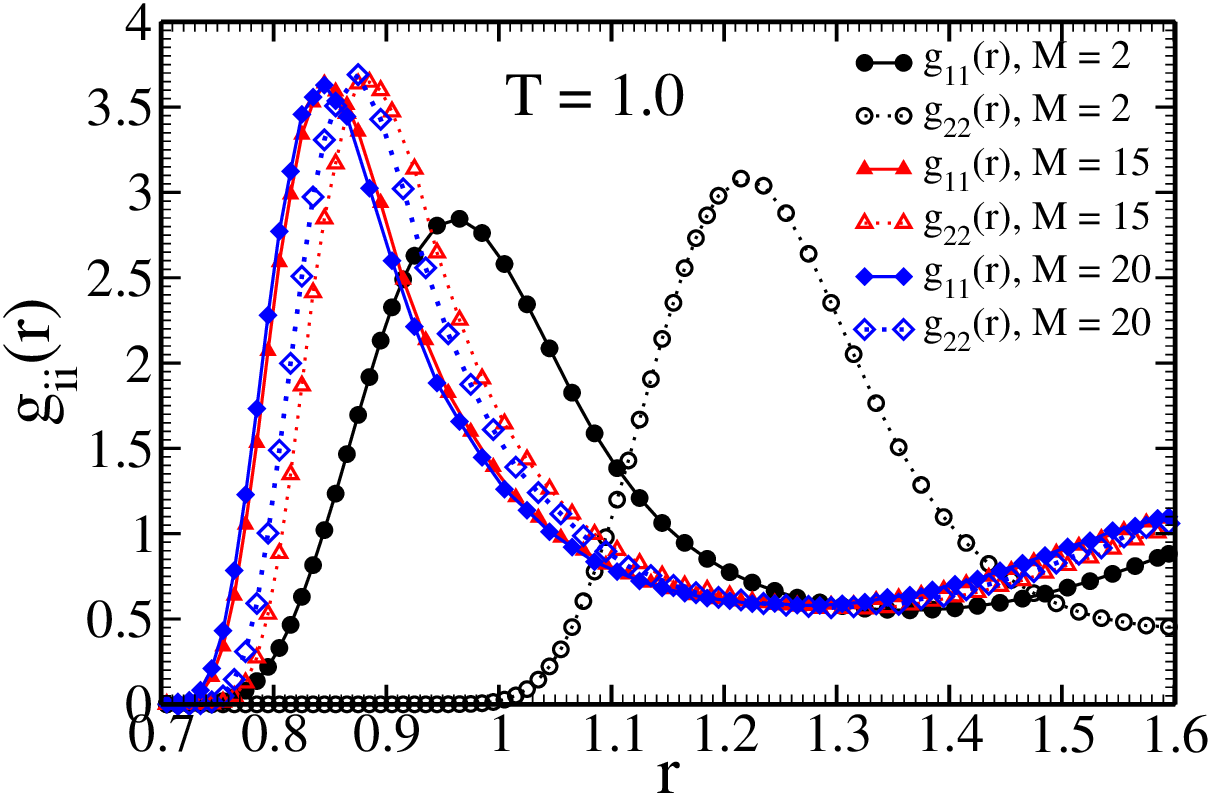}
    \caption{\emph{The partial radial distribution function for the first two species for different values of $M$. The particles are interacting via WCA potential and the polydispersity of the system is described by $P_{1}(\sigma)$ with $15\%$ PDI.}}
    \label{gr_WCA}
\end{figure}

To understand the origin of this lower tolerance of the WCA system in Fig.\ref{gr_WCA} we plot for the WCA system the partial rdfs of the first two species for different values of $M$ and find that compared to the IPL system we need to go to higher values of $M$ to observe a good overlap between the two rdfs. Similar to that found for the IPL system, at $M=M_{0}$ the rdf peaks almost overlap. Note that compared to the WCA potential the IPL potential is softer and has a longer range. Thus compared to the WCA system the IPL system has a flatter rdf and a larger overlap of the rdfs of two consecutive species. This definitely explains why compared to the IPL system the $M_{0}$ is higher for the WCA system.

\begin{figure}
    \centering
    \includegraphics[width=.45\textwidth]{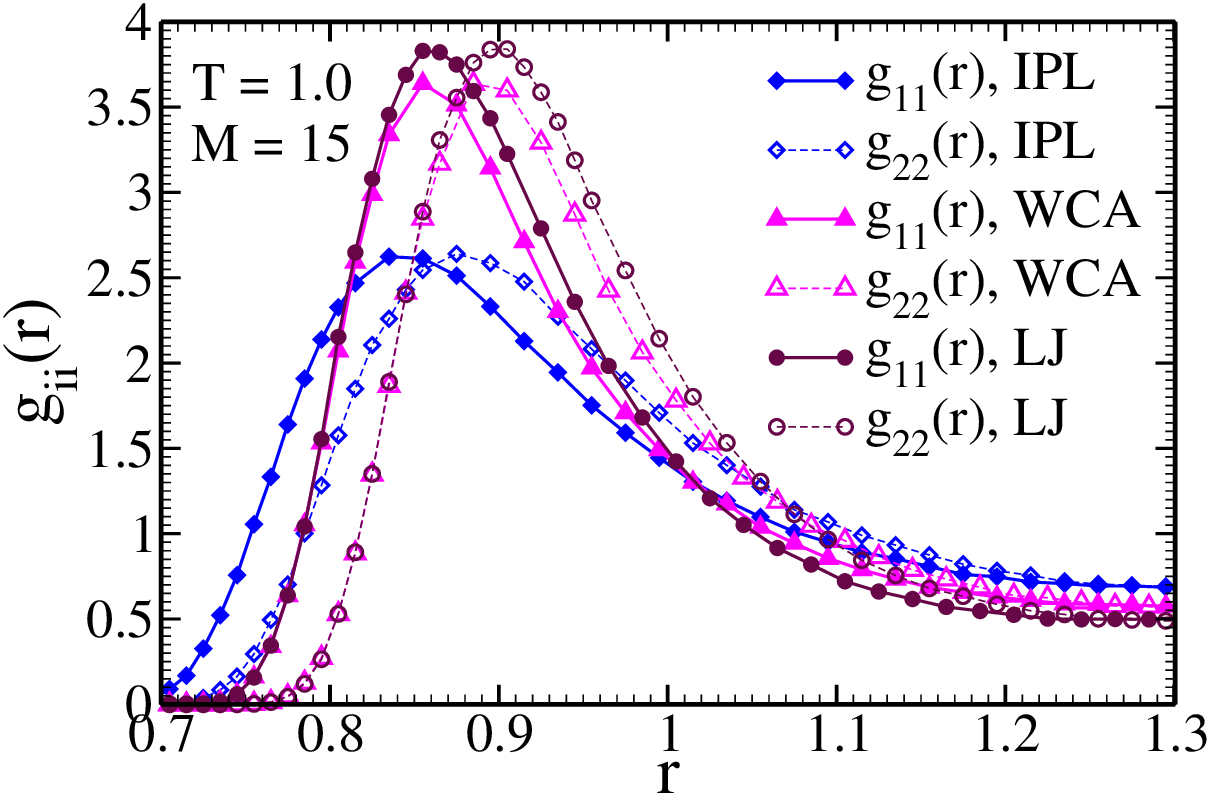}
    \caption{\emph{The partial radial distribution functions for the first two species for IPL, WCL and LJ potentials for $M=15$. The size distribution of the particles is given by $P_{1}(\sigma)$ with $15\%$ polydispersity.}}
    \label{gr_diff_potential_M_15}
\end{figure}

In Fig.\ref{gr_diff_potential_M_15} we compare the rdf values for the first two species of the IPL, WCA and LJ systems, for M = 15. Note that for the IPL and the LJ systems  $M_{0} \le 15$ and for the WCA system $M_{0}>15$.
However, compared to the WCA and IPL systems, the partial rdfs for the LJ system are more sharply peaked. This does not explain why the $M_{0}$ for the LJ system is smaller than the WCA system. Note that the structure along with the interaction potential describes both the potential energy and also the two body entropy.  In $S_{2}$ the leading term is -$g(r) \ln g(r) \simeq g(r) W(r)$ where $W(r)=-\ln g(r)$ can be considered as the effective potential of the system. Thus along with the rdf the interaction potential also plays a role in determining these thermodynamic quantities. The range of the LJ potential is much larger compared to the IPL and WCA potentials. Moreover, the attractive part of the potential which provides a dominant contribution also varies much more smoothly compared to the WCA and IPL potentials. It appears that this slow variation of the potential increases the tolerance of thermodynamic quantities w.r.t the particle size which leads to a smaller $M_{0}$ value.

We will like to mention that in this work while working with the LJ system we only vary the size of the particles while keeping the interaction energy constant. This choice is quite similar to that used in earlier studies of model polydisperse systems \cite{PDI_bagchi_sneha_sarika,Poole_PDI,Tanaka}. However, this choice of system is not consistent with the van der Waals attraction dependence on particle volume. Thus the system can have some unusual structures like that of clustering of smaller particles observed earlier \cite{Tanaka}

\subsection{System size dependence}
\begin{figure}
    \centering
    \includegraphics[width=.45\textwidth]{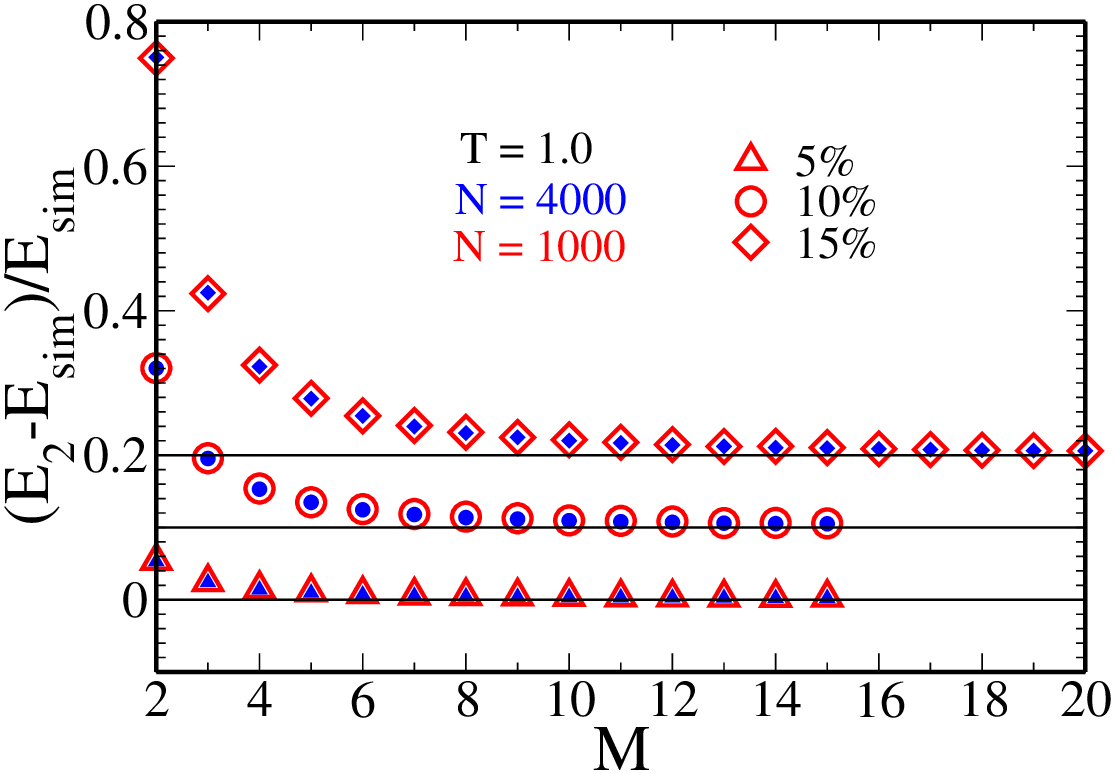}
    \caption{\emph{System size dependence of relative error calculation between $E_{sim}$ and $E_{2}$. ($E_{2}-E_{sim})/E_{sim}$ plotted as a function of $M$ for different PDIs for  N = 1000 (red open symbols) and N = 4000 (blue filled symbols). For better visualization, we have shifted the y-axis of the 10\% PDI plot by 0.1 and 15\% PDI plot by 0.2. The horizontal lines signify the corresponding zero values.}}
    \label{E_4k_particle}
\end{figure}

\begin{figure}
    \centering
    \includegraphics[width=.45\textwidth]{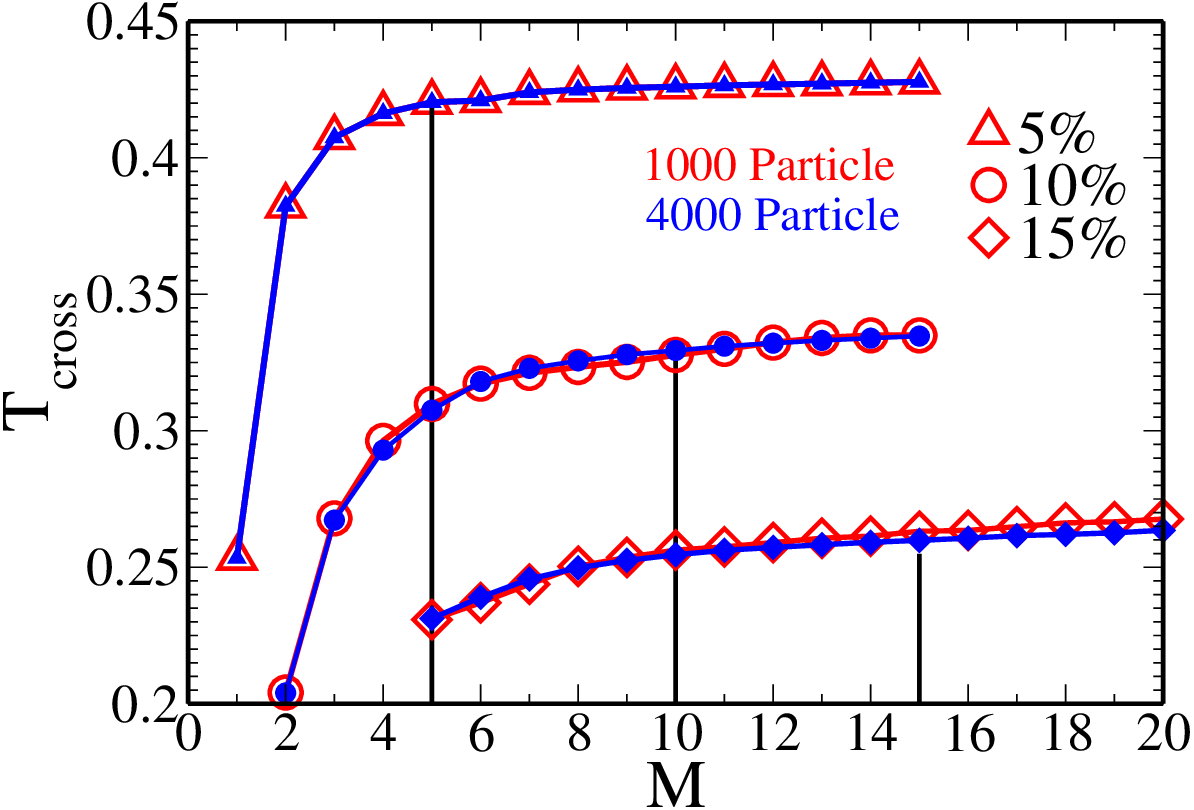}
    \caption{\emph{The system size dependence of $M_0$. $T_{cross}$ vs $M$ plot for different PDIs for systems where particles are interacting via IPL potential and the particle size distribution are given by $P_{1}(\sigma)$. The open red symbols are for N=1000 and the filled blue symbols are for N=4000. The vertical lines give the value of $M_{0}$ obtained from energy criteria. $M_{0}$ from energy is independent of system size and for systems with higher PDI the $T_{cross}$ saturates better for higher system size.}}
    \label{T_cross_system_size}
\end{figure}

\begin{figure}
    \centering
    \includegraphics[width=.45\textwidth]{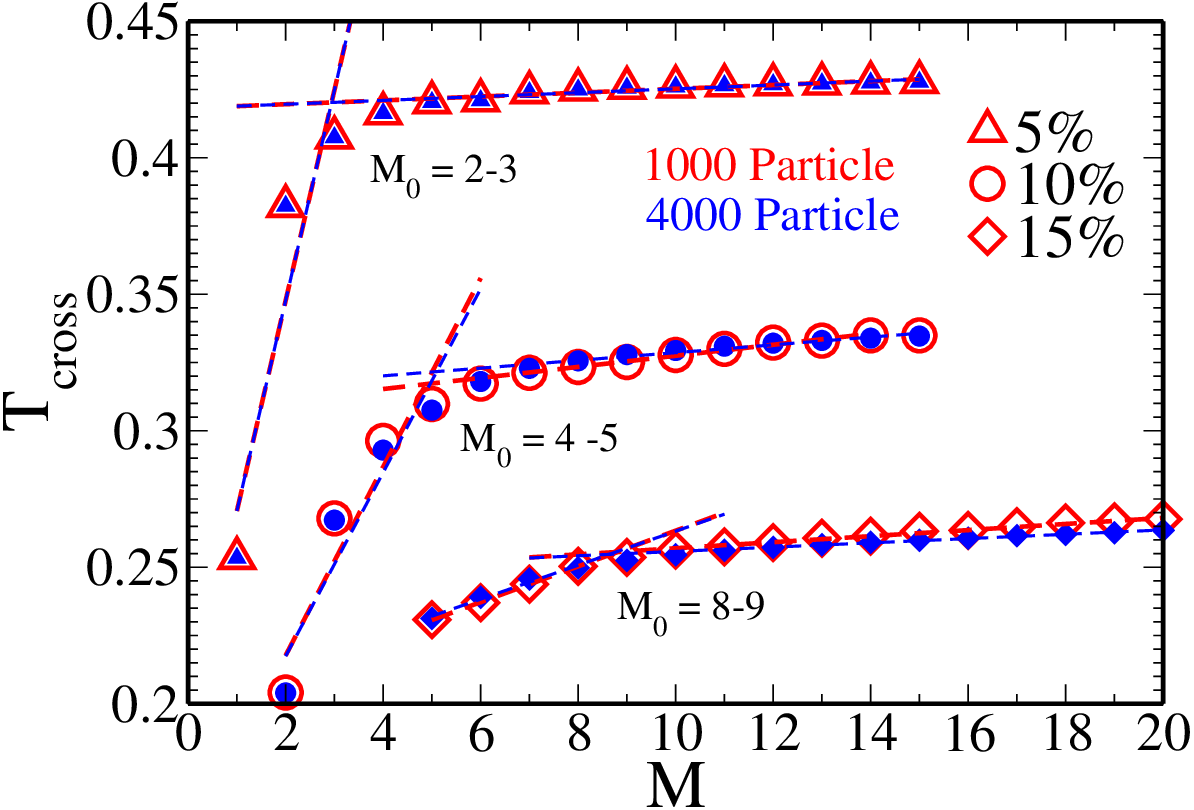}
    \caption{\emph{Alternate definition of $M_{0}$. $T_{cross}$ vs M has two different regimes, low M shows growth and high M shows a near saturation. The two different regimes are fitted to two straight lines and the M value where these lines cross is identified as $M_{0}$. The open red symbols are for N=1000 and the filled blue symbols are for N=4000. The $M_{0}$ values thus obtained are smaller than that obtained from energy criteria and similar to that obtained by Ozawa and Berthier \cite{ozawa}}}
    \label{T_cross_fit_system_size}
\end{figure}

Note that for finite number of particles ($N$) in the system, we can still describe the $N*N$ partial structure factors. However in the thermodynamic limit when $N\rightarrow \infty$ this becomes ill-defined. Thus for larger systems describing the system in terms of pseudo species becomes even more important. In this section, we study the system size dependence of $M_{0}$. For this study, we take the system where particle size distribution is given by $P_{1}(\sigma)$ and the particles interact via IPL potential. We study three systems at $5\%$, $10\%$ and $15\%$ PDI. In Fig.\ref{E_4k_particle} we plot the $\frac{(E_{2}-E_{sim})}{E_{sim}}$ for systems with 1000 and 4000 particles. We find that the relative error in energy is independent of the system size and so is $M_0$. In Fig.\ref{T_cross_system_size} we plot the $T_{cross}$ values for the two different system sizes. We find that for $5\%$ and $10\%$ PDI they overlap and for $15\%$ PDI the bigger system size predicts a better saturation of $T_{cross}$. This is because we now have a larger number of particles in each species giving rise to better averaging. Thus we can say that $M_{0}$ is independent of system size.

\subsection{Comparison with earlier predictions}
 
Next, we compare our results with that of Ozawa and Berthier (OB) \cite{ozawa}. The goal of both studies is to find an effective number of pseudo species that can describe a polydisperse system. However, the methodologies are quite different. We work with the partial rdf of the liquid and use them to calculate the excess entropy and total energy near the onset temperature. Ozawa and Berthier used the information of the vibrational entropy and the inherent structure properties and the study was performed below the MCT transition temperature. They divided the system into M species and then swapped particles within a species. After N such swaps, they minimized the system and obtained the mean square displacement (MSD) between the original equilibrium configuration and swapped configuration. The MSD as a function of M initially decreased with increasing M and showed saturation at high values of M. They fitted these two regimes to two different power laws, and the intersection point of the power laws determined the value of $M_{0}$.
In their study they have calculated the value of $M_{0}$ for an IPL potential system with $P_{1}(\sigma)$ distribution at 23$\%$ PDI. They then extrapolated the value of $M_{0}$ for smaller PDIs. For 10$\%$ PDI they predicted a value of $M_{0}= 4-5$ and for 5$\%$ PDI, $M_{0}=2-3$.
We can do a similar exercise with $T_{cross}$. In Fig.\ref{T_cross_fit_system_size} we show that $T_{cross}$ also shows two different regimes. We fit the two different regimes to two straight lines and the point where they cross is, $M_{0}=2-3$ for 5\% PDI, $M_{0}=4-5$ for 10\% PDI, and $M_{0}=8-9$ for 15\% PDI. Interestingly these numbers are surprisingly the same as that extrapolated by Ozawa and Berthier \cite{ozawa} although the two methodologies are completely different. However, these values of $M_{0}$ are lower than our earlier prediction which was made by looking at the saturation point of $T_{cross}$. In the OB study if they define the $M_{0}$ at the value where their MSD becomes independent of $M$ then they too will have a higher value of $M_{0}$.

We next compare our predictions with an earlier work which involved the study of the dynamics  \cite{voigtmann_polydis_mct}. 
As discussed in the Introduction, Weysser {\it et al} studied the effect of polydispersity on the dynamics \cite{voigtmann_polydis_mct}. They studied a system with constant polydispersity where $\Delta \sigma=0.2$. According to their study, the dynamics can be well explained when the system is divided into 5 pseudo species and thus $\Delta \sigma_{0}=0.04$ falls in a similar range as predicted here and so is $M_{0}$.

At this point, we cannot comment on which will be a better choice of $M_{0}$, the value at which $T_{cross}$ saturates or the value at which two different regimes intersect. When we compare our result with the study using the dynamics \cite{voigtmann_polydis_mct} it appears that the former which leads to higher values of $M_{0}$ is a better choice whereas if we compare with OB study then the latter seems to be a better choice. It is possible that the dynamics is more sensitive to change in $M_{0}$. This suggests that further studies are required to narrow down the value of $M_{0}$. One such option will be to see how the configurational entropy for different values of $M_{0}$ correlates with the dynamics.

\section{conclusion}
In this work we attempt to develop a framework to describe the structure of systems with continuous polydispersity. The study involves systems where the polydispersity is described in terms of different distributions (constant volume fraction and Gaussian) and the degree of polydispersity is varied. We also study the effect of the interaction potential. 

We exploit the fact that the potential energy and the pair excess entropy can be described in terms of the partial radial distribution functions. First, we describe the system in terms of pseudo species. In the case of potential energy, we obtain the minimum number of pseudo species, $M_{0}$ required to match the energy obtained from the partial rdf with that obtained from the simulation. For the entropy part, since the excess entropy and pair excess entropy are never equal, we calculate the temperature where they cross each other. Our earlier study has shown that this $T_{cross}$ is an estimate of the onset temperature of supercooled liquids\cite{onset_crosspoint}. We show that for a  smaller number of species, the $T_{cross}$ varies with $M$ and as a function of species this temperature shows a saturation, suggesting a saturation of the pair excess entropy w.r.t $M$. This gives us a second estimation of $M_{0}$ which we find is similar to that obtained from the potential energy. 

Our study reveals that for a given system, it is possible to define a parameter $\Delta \sigma_{0}$ which gives the limiting width of the size distribution that can be treated as a monodisperse system. This limiting value primarily depends on the interaction potential. The softer the interaction potential  the larger is the value of $\Delta \sigma_{0}$. Depending on the type of distribution this limiting width $\Delta \sigma_{0}$ translates into different values of PDI. 

For a 1$\%$ PDI system with constant volume fraction distribution, $\Delta\sigma=0.036$ and with Gaussian distribution,$\Delta\sigma=0.06$. When we compare these values with $\Delta\sigma_{0}$, we can say that polydispersity greater than $1\%$ when treated as an effective monodisperse system will not provide us with the correct structure of the liquid. This implies that when the effective  one component structure is used to study the influence of polydispersity on some property, we have to be careful in decoupling the effect of this artificial softening of the structure from the actual effect of the polydispersity. Note that $M_{0}$ and $\Delta \sigma_{0}$ are independent of the system size. This makes this pseudo  neighbour description of a system more useful for larger systems.

{\bf Dedication}\\
This article is dedicated to the memory of Prof. Charusita Chakravarty a mentor and a friend. She not only made a huge contribution in Physical Chemistry, but also was passionate about the upliftment of women scientists. Had she been around she would have definitely felt very happy about this special issue and contributed to it. Her absence has created a void in the field. \\

{\bf Acknowledgment}\\
S.~M.~B thanks C. Chakravarty, C.~ Dasgupta, S.~ Sastry, and B.~ Bagchi for discussion. P.~P. and U.~K.~N thanks CSIR, for the research fellowships. M.~K.~N thanks CSIR for funding, S.~M.~B. thanks SERB for funding.\\[4mm]

{\bf Availability of Data}\\
The data that support the findings of this study are available from the corresponding author upon reasonable request.
 
\section{References}


\end{document}